\documentclass[12pt]{article}
\usepackage[dvips]{graphicx}
\usepackage{amssymb}
\usepackage{amsmath}

\def\half{{\textstyle{\frac12}}}

\def\beq{\begin{equation}}\def\eeq{\end{equation}}
\def\bea{\begin{eqnarray}}\def\eea{\end{eqnarray}}

\begin{document}

\title{Realistic interpretation of Grassmann variables}
\author{Roman Sverdlov,
\\Department of Physics, University of Mississippi} 

\date{March 26, 2015}
\maketitle

\begin{abstract}
The goal of this paper is to define the Grassmann integral in terms of a limit of a sum around a well-defined contour so that Grassmann numbers gain geometric meaning rather than symbols. The unusual rescaling properties of the integration of an exponential is due to the fact that the integral attains the known values only over a specific set of contours and not over their rescaled versions. Such contours live in infinite dimensional space and their sides are infinitesimal, and they make infinitely many turns. Finally, two different products are used: anticommutting wedge product and a Clifford dot product (the wedge product is used in the finite part of the integral and the Clifford dot product is used between the finite and infinitesimal parts). The integrals of non-analytic functions will become well-defined, although their specific value is unknown due to the various hidden parameters. 
\end{abstract}

\section{Introduction}

In light of the fact that fermions anticommute, in quantum field theory anticommutting \emph{Grassmann variables} are used to model fermionic path integrals, which satisfy 
\beq \theta_1 \theta_2 = - \theta_2 \theta_1 \eeq
In light of the fact that the square of anticommutting number is zero, 
\beq \theta^2 =0 \eeq
all analytic functions become linear. For example,
\beq e^{k \theta} = 1 + k \theta \label{ConventionalExponential} \eeq
The integral of a general such linear function is defined
\beq \int d \theta \; (a+ b \theta) = b \eeq
and, therefore, in light of Eq \ref{ConventionalExponential},
\beq \int d \theta e^{k \theta} = k \label{ExponentialDoesntAnticommute} \eeq
Conventionally, Grassmann integration is viewed as merely an algebraic operation as opposed to the limit of the sum, for two reasons:

1. The properties of the integral contradict the expected ones. For example, the integral over $e^{k \theta}$ is proportional to $k$ rather than $k^{-1}$, the integral over the constant is zero and the integral over the odd function is not, and so forth. 

2. Even though the product of two anticommutting numbers is commutting, it is still not a real number: after all, $(\theta_1 \theta_2)^2 =0$. So how can the sum of such products -- in particular, the sum of $d \theta \; \theta$ -- possibly be real? 

We address both of those questions by replacing
\beq \int d \theta \; f (\theta) g (\theta) \eeq
with
\beq \int_{\Gamma} d \theta \cdot (f (\theta) \wedge g (\theta)) \eeq
where

1) $\Gamma$ is a \emph{carefully selected} contour. Thus, we have multi-dimensional space, while $\theta$ is being confined to the contour living in that space. Furthermore, we claim that the integration results match the conventional ones only over a particular set of contours, not all of them. 

2) The dot-product is distinct from the wedge-product; in particular, $\theta_1 \cdot \theta_2 = \theta_1 \wedge \theta_2 + \lambda (\theta_1, \theta_2)$, where $\lambda$ is a symmetric, bi-linear, real valued function, such that $\lambda (\theta, \theta) = \vert \theta \vert^2$ and $\lambda (\theta_1, \theta_2) = 0$ if $\theta_1$ is orthogonal to $\theta_2$.

If we claim that the integral obeys expected properties \emph{only} over said $\Gamma$ \emph{as opposed to} any other contour, we can then claim that 
\beq \int_{\Gamma} d \theta \cdot e^{k \theta} = \frac{1}{k} \int_{k \Gamma} d \theta \cdot e^{\theta} = \frac{1}{k} k^2 = k \label{WrongContourIntro} \eeq
Thus, we appealed to the fact that the integral over $k \Gamma$ of $e^{\theta}$ returns $k^2$ instead of $1$. That is because we never said that the integral returns $1$ over \emph{all} contours. We \emph{only} said that it returns $1$ over \emph{some particular contour}; therefore, we are still free to say that it returns $k^2$ over the rescaled version of that contour, which removes the contradiction. 

As far as the second question, how can the integral return the real number, as long as we have 
\beq \theta_1 \cdot \theta_2 = \theta_1 \wedge \theta_2 + \lambda (\theta_1, \theta_2) \eeq
we can always try to design the sum of \emph{dot-}products in such a way that the wedge-product terms cancel out while the $\lambda$-terms add up to whatever real number we would like to get. This is accomplished by designing the contour in the appropriate way.  

In a \emph{typical} case, the presence of the dot-product is irrelevent when it comes to the definition of functions \emph{under} the integral for the simple reason that said functions happened to be defined in terms of the wedge product alone, which is what we mean by the word \emph{analytic}: for example, 
\beq e^{f (\theta)} = 1 + f (\theta) + \frac{1}{2} f (\theta) \wedge f (\theta) + \frac{1}{6} f (\theta) \wedge f (\theta) \wedge f (\theta) + \cdots \label{FiniteWedge} \eeq
At the same time, one of the implications of this paper is that the theory can be extended to non-analytic functions. One possibility of non-analytic function is replacing wedge with dot. In fact, in Chapter \ref{NonAnalyticExponential} we have computted what would happen if we were to do just that. But, as we show, the conventional results are only reproduced if we use wedge in the finite part, and the sole purpose of using dot in finite part might be if we want to go outside of conventional framework (theory of quantum measurement and so forth). At the same time, when it comes to differential part, yes we do use dot, even in conventional case. 

\section{Definition of products}

Before we proceed any further, let us define the products we just talked about. We start from an infinite dimensional space, with the unit vector along the dimension $k$ being $e_k$. We then assume that the general element takes the form 
\beq G=  g + \sum_{k=1}^{\infty} g_k e_k + \sum_{k<l} g_{kl} e_k \wedge e_l + \sum_{i<j<k} g_{ijk} e_i \wedge e_j \wedge e_k + \cdots = \nonumber \eeq
 \beq = g + \sum_{l=1}^{\infty} \sum_{k_1 < \cdots < k_l} g_{k_1, \cdots, k_l} e_{k_1} \wedge \cdots \wedge e_{k_l} \eeq 
where $g$, $g_k$, $g_{kl}$, $g_{ijk}$, and so forth, are real numbers, 
\beq g_{a_1 \cdots a_k} \in \mathbb{R} \eeq
and, therefore, commute
\beq g_{a_1 \cdots a_k}  g_{b_1 \cdots b_l} = g_{b_1 \cdots b_l} g_{a_1 \cdots a_k} \eeq
The anticommutting part comes from unit vectors $e$:
\beq e_k \wedge e_l = - e_l \wedge e_k \eeq
If $k \neq l$, then the two products agree:
\beq e_k \cdot e_l = e_k \wedge e_l \; , \; k \neq l \label{DotWedgeAgree} \eeq
Their disagreement comes from where $k= l$:
\beq e_k \cdot e_k =1 \; , \; e_k \wedge e_k = 0 \label{DotWedgeDisagree} \eeq 
Finally, these products agree when it comes to multiplication by a real number:
\beq r \in \mathbb{R} \Longrightarrow  r \wedge G = G \wedge r = r \cdot G = G \cdot r = rG \eeq
where $rG$ without a dot or a wedge stands for vector space scalar multiplication. We then generalize Eq \ref{DotWedgeAgree} as
\beq [\forall i \neq j (a_k \neq a_j)] \Longrightarrow e_{a_1} \cdot e_{a_2} \cdot \cdots \cdot e_{a_{n-1}} \cdot e_{a_n} = e_{a_1} \wedge e_{a_2} \wedge\cdots \wedge e_{a_{n-1}} \wedge e_{a_n} \eeq
Once again, the assumption $a_k \neq a_j$ is crucial. For example, if we were to have $a_1 = a_2$ then, per Eq \ref{DotWedgeDisagree}, we would have had 
\beq (a_1 = a_2 \; , \; e_k \neq e_l, 2 \leq k < l) \Longrightarrow \nonumber \eeq
\beq \Longrightarrow e_{a_1} \cdot e_{a_2} \cdot \cdots \cdot e_{a_{n-1}} \cdot e_{a_n} = 1 \cdot e_{a_3} \cdot \cdots \cdot e_{a_{n-1}} \cdot e_{a_n} = \nonumber \eeq
\beq = e_3 \cdot \cdots \cdot e_{a_{n-1}} \cdot e_{a_n} = e_3 \wedge \cdots \wedge e_{a_n} \label{DotShortened} \eeq
In other words we would have $e_{a_3} \wedge \cdots \wedge e_{a_n}$ in $a_1 =a_2$ case, in contrast to $e_{a_1} \wedge \cdots \wedge e_{a_n}$ in $a_1 \neq a_2$ case. This should also be contrasted with the wedge product where we have 
\beq a_1 = a_2 \Longrightarrow e_{a_1} \wedge \cdots \wedge e_{a_n} = 0 \eeq
which is not true for the dot product:
\beq (a_1 = a_2 \; , \; e_k \neq e_l, 2 \leq k < l) \Longrightarrow e_{a_1} \cdot \cdots \cdot e_{a_n} = e_{a_3} \wedge \cdots \wedge e_{a_n} \neq 0 \eeq
 Notably, in Eq \ref{DotShortened} we have also used associativity, as evident from the first equal sign below:
\beq a_1 = a_2 \Longrightarrow e_{a_1} \cdot (e_{a_2} \cdot e_{a_3} \cdot \cdots \cdot e_{a_{n-1}} \cdot e_{a_n}) = (e_{a_1} \cdot e_{a_2}) \cdot (e_{a_3} \cdot \cdots \cdot e_{a_{n-1}} \cdot e_{a_n}) =  \nonumber \eeq
\beq = 1 \cdot (e_{a_3} \cdot \cdots \cdot e_{a_{n-1}} \cdot e_{a_n}) \eeq
It turns out that associativity is actually quite difficult to prove. But, for the purposes of the physics paper, we will just assume associativity holds based on the intuition we have derived from $\gamma$-matrices and so forth. Let me now give a few other examples to illustrate how typical calculation works:
\beq (e_1 \wedge e_3) \cdot (e_2 \wedge e_3) = - (e_1 \wedge e_3) \cdot (e_3 \wedge e_2) = - (e_1 \cdot e_3) \cdot (e_3 \cdot e_2) = \nonumber \eeq
\beq =  - e_1 \cdot (e_3 \cdot e_3) \cdot e_2 = - e_1 \cdot 1 \cdot e_2 = - e_1 \cdot e_2 = - e_1 \wedge e_2 \eeq
and, on the other hand, 
\beq (e_1 \wedge e_3) \cdot (e_2 \wedge e_4) = (e_1 \cdot e_3) \cdot (e_2 \cdot e_4) = e_1 \wedge e_3 \wedge e_2 \wedge e_4 = - e_1 \wedge e_2 \wedge e_3 \wedge e_4 \eeq
Notice that the second calculation could have been done differently: 
\beq (e_1 \wedge e_3) \cdot (e_2 \wedge e_4) = (e_1 \cdot e_3) \cdot (e_2 \cdot e_4) = e_1 \cdot (e_3 \cdot e_2) \cdot e_4 = e_1 \cdot (e_3 \wedge e_2) \cdot e_4 = \nonumber \eeq
\beq = - e_1 \cdot (e_2 \wedge e_3) \cdot e_4 = e_1 \cdot (e_2 \cdot e_3) \cdot e_4 = - e_1 \wedge e_2 \wedge e_3 \wedge e_4 \eeq
Notice that in both cases we got the same answer. Once again, actual proof that the answers will always match is quite difficult, but for the sake of physics paper we will simply trust that that's the case. Finally, to give an example where the minus sign does not appear in the final answer, 
\beq (e_1 \wedge e_3) \cdot (e_1 \wedge e_2 \wedge e_3) = (e_1 \wedge e_3) \cdot (e_3 \wedge e_1 \wedge e_2) = (e_1 \cdot e_3) \cdot (e_3 \cdot e_1 \cdot e_2) = \nonumber \eeq
\beq = (e_1 \cdot (e_3 \cdot e_3) \cdot e_1) \cdot e_2 = (e_1 \cdot 1 \cdot e_1) \cdot e_2 = (e_1 \cdot e_1) \cdot e_2 = 1 \cdot e_2 = e_2 \eeq 
and, on the other hand, the minus sign again appears in 
\beq (e_1 \wedge e_2) \cdot (e_1 \wedge e_2 \wedge e_3) = - (e_2 \wedge e_1) \cdot (e_1 \wedge e_2 \wedge e_3) = - (e_2 \cdot e_1) \cdot (e_1 \cdot e_2 \cdot e_3) = \nonumber \eeq
\beq = - (e_2 \cdot (e_1 \cdot e_1) \cdot e_2) \cdot e_3 = - (e_2 \cdot 1 \cdot e_2) \cdot e_3 = - (e_2 \cdot e_2) \cdot e_3 = - 1 \cdot e_3 = - e_3 \eeq

\section{Definition of contours and single variable integrals}

Now that we have defined the products, we are ready to go on to the next step and define the contours that would produce the desired outcomes of integration. For any $a \in \mathbb{R}$ and $d \in \mathbb{N}$, let us define the contour $\Gamma_{d,a} (t)$ in the following way: 
\beq \Gamma_{d,a} (t) =  \left\{ \begin{array}{lll}
         0 & t \leq 0 \\
        a (e_1 + \cdots + e_{k-1}) + u e_k & 0 \leq k \leq d-1, t = k+u, 0 \leq u \leq a \\
        a (e_1 + \cdots + e_n) & t \geq an   \end{array} \right. \eeq
It is easy to see that 
\beq \int_{\Gamma_{d,a}} d \theta = a (e_1 + \cdots + e_d) \eeq
and, with slightly more complicated calculation, one can show that
\beq \int_{\Gamma_{d,a}} d \theta \cdot \theta = \sum_{k=1}^d \int_0^a [(d u \;  e_k) \cdot (a (e_1 + \cdots + e_{k-1}) + ue_k)] = \nonumber \eeq
\beq = \sum_{k=1}^d \bigg( (e_k \cdot e_k) \int_0^a du \; u + a \sum_{l=1}^{k-1} \bigg( (e_k \cdot e_l) \int_0^a du \bigg) \bigg) = \nonumber \eeq 
\beq = \sum_{k=1}^d \bigg( \frac{a^2}{2} e_k \cdot e_k + a^2 \sum_{l=1}^{k-1} e_k \cdot e_l \bigg)  = \frac{a^2}{2} \sum_{k=1}^d 1 + a^2 \sum_{1 \leq l < k \leq d} e_k \wedge e_l = \nonumber \eeq
\beq = \frac{da^2}{2} - a^2 \sum_{1 \leq l < k \leq d} e_l \wedge e_k \eeq 
If we now set 
\beq a = \sqrt{\frac{2}{d}} \label{PreferredChoice} \eeq 
we obtain 
\beq \int_{\Gamma_{d, \sqrt{2/d}}} d \theta = \sqrt{\frac{2}{d}} (e_1 + \cdots + e_d) \label{AlmostZero1Var} \eeq
\beq \int_{\Gamma_{d, \sqrt{2/d}}} d \theta \cdot \theta = 1 - \frac{2}{d} \sum_{1 \leq l < k \leq d} e_l \wedge e_k  \eeq 
Now, \emph{if} we were to find a way of getting rid of the non-real parts, this \emph{would} leave us with the $0$ and $1$ that we "want". Whether or not we can do that depends on how we define our metric and limit procedure. On the one hand, in the limit of $d \rightarrow \infty$, each \emph{individual} non-real component is small:
\beq \lim_{d \rightarrow \infty} \sqrt{\frac{2}{d}} = 0 \; , \; \lim_{d \rightarrow \infty} \bigg(- \frac{2}{d} \bigg) = 0 \eeq
 on the other hand, the Eucledian norm of the sum of \emph{all of them} is not: 
\beq \sqrt{\frac{2}{d}} \sqrt{\sum_{k=1}^d 1^2} = \sqrt{\frac{2}{d}} \sqrt{d} = \sqrt{2} \label{Root2}  \eeq
\beq \lim_{d \rightarrow \infty} \bigg( \frac{2}{d} \sqrt{\sum_{1 \leq l < k \leq d} 1^2} \bigg)  = \lim_{d \rightarrow \infty} \bigg( \frac{2}{d} \sqrt{\frac{d(d-1)}{2}} \bigg) = 1 \label{Bad1}  \eeq
In order to avoid these issues, we borrow the definition of sup-norm and write 
\beq \bigg\vert g + \sum_{l=1}^{\infty} \sum_{k_1 < \cdots < k_l} g_{k_1 \cdots k_l} e_{k_1} \wedge \cdots \wedge e_{k_l} \bigg\vert_{\rm max}  = \max (\{ g \} \cup \{g_{k_1, \cdots, k_l} \vert l \in \mathbb{N} \}) \label{NormMax} \eeq
We then define $\lim^{max}$ with respect to the above max-norm as follows: 
\beq \lim^{max}_{n \rightarrow \infty} \bigg( g_n + \sum_{l=1}^{\infty} \sum_{k_1 < \cdots < k_l} g_{n; k_1, \cdots, k_l} e_{k_1} \wedge \cdots \wedge e_{k_l} \bigg)  =  h + \sum_{l=1}^{\infty} \sum_{k_1 < \cdots < k_l} h_{k_1, \cdots, k_l} e_{k_1} \wedge \cdots \wedge e_{k_l} \Longrightarrow \nonumber \eeq
\beq \Longrightarrow \forall \epsilon >0 \exists N \in \mathbb{N} \forall n > N ((\vert g_n- h \vert < \epsilon)  \wedge \forall l \in \mathbb{N} \vert g_{n; k_1, \cdots, k_l} - h_{k_1, \cdots, k_l} \vert < \epsilon)  \label{LimMax} \eeq
Notice that Eq \ref{LimMax} is only true if the norm is defined per Eq \ref{NormMax}. As one can see from Eq \ref{Root2} and \ref{Bad1}, under the Eucledian norm the Eq \ref{LimMax} will no longer hold. However, while norm-max is not rotationally invariant, the lim-max is -- \emph{provided that by rotation we mean a mixture of only finitely many coordinates} which, henceforth, we will call "finite coordinate rotation". That is due to the fact that the same topology is being generated by many different norms. Clearly, norm-max change under finite coordinate rotations, yet topology-max remains the same, and so does lim-max. In any case, the important result is that 
\beq \lim^{max}_{d \rightarrow \infty} \int_{\Gamma_{d, \sqrt{2/d}}} d \theta = 0 \label{OneVariableConstant} \eeq
\beq \lim^{max}_{d \rightarrow \infty} \int_{\Gamma_{d, \sqrt{2/d}}} d \theta \cdot \theta = 1 \label{RightContourDemo} \eeq 
The situation with the wrong choice of contours seen in Eq \ref{WrongContourIntro} can be reproduced per
\beq \lim^{max}_{d \rightarrow \infty} \int_{\Gamma_{d, k \sqrt{2/d}}} d \theta \cdot \theta = k^2 \label{WrongContourDemo} \eeq 
where the only difference between the left hand sides of Eq \ref{RightContourDemo} and Eq \ref{WrongContourDemo} is $\Gamma_{d, \sqrt{2/d}}$ being used in the first case and $\Gamma_{d, k \sqrt{2/d}}$ in the second. 

\section{Sign conventions in multiple integrals} \label{SignSection}

Before we proceed to investigate multiple integrals, it is important that we are on the same page when it comes to signs -- although this is merely a conventional issue that is a lot less important than the other things we talk about. Traditionally, it is assumed that 
\beq \int d \theta_1 d \theta_2 \; \theta_1 \theta_2 = - \int d \theta_1 d \theta_2 \; \theta_2 \theta_1 = +1 \eeq
However, from a logical point of view we would expect 
\beq \int d \theta_1 d \theta_2 \; \theta_1 \theta_2 = - \int d \theta_1 d \theta_2 \; \theta_2 \theta_1 = - \int \bigg[ d \theta_1 \bigg( \int d \theta_2 \; \theta_2 \bigg) \theta_1 \bigg] = \nonumber \eeq
\beq =  - \int d \theta_1 \; 1 \theta_1 = - \int d \theta_1 \theta_1 = -1 \eeq 
The way we resolve the two is by claiming that, whenever there is a product $*$, there is a corresponding \emph{inverted product} $\overline{*}$ defined as 
\beq a \overline{*} b = - a* b \eeq
As long as $*$ is associative, $\overline{*}$ is associative as well, as evident from the following: 
\beq a \overline{*} (b \overline{*} c) = a \overline{*} (- b*c) = - a* (-b*c) = a*(b*c) \eeq
\beq (a \overline{*} b) \overline{*} c = (- a*b) \overline{*} c = - ((-a*b)*c) = (a*b)*c \eeq 
Therefore, we have 
\beq \int (d \theta_1 * d \theta_2) \cdot (\theta_1 \wedge \theta_2) = -1 \Longleftrightarrow  \int (d \theta_1 \overline{*} d \theta_2) \cdot (\theta_1 \wedge \theta_2) =+1 \eeq
The general integral that would lead to conventional signs takes the form
\beq \int (d \theta_1 \overline{*} d \theta_2 * d \theta_3 \overline{*} d \theta_4 *\cdots * d \theta_{2k-1} \overline{*} d \theta_{2k} * \cdots * d \theta_{2n-1} \overline{*} d \theta_{2n}) \cdot (\theta_1 \wedge \cdots \wedge \theta_{2n})=1  \eeq 
\beq \int (d \theta_1 \overline{*} d \theta_2 * d \theta_3 \overline{*} d \theta_4 *\cdots * d \theta_{2k-1} \overline{*} d \theta_{2k} * \cdots \overline{*} d \theta_{2n} * d \theta_{2n+1}) \cdot (\theta_1 \wedge \cdots \wedge \theta_{2n+1})=1  \eeq 
whereas the form that leads to sign convention we prefer to use throughout this paper is 
\beq \int (d \theta_1 * \cdots * d \theta_n) \cdot (\theta_1 \wedge \cdots \wedge \theta_n) = (-1)^{n(n+1)/2} \eeq
where the shortened $d\theta_1 * \cdots * d\theta_n$ and $\theta_1 \wedge \cdots \wedge \theta_n$ indicate that neither $\overline{*}$ nor $\overline{\wedge}$ are used. All three of the above formulae are simultaneously correct, so the issue of sign convention is merely an issue as to which of those formulae we prefer to use. 

\section{Double integral of $f (\theta_1, \theta_2)=1$ and importance of order of limits}

Let us now turn to multiple integrals. First of all, we have to be careful as to how we take the integral, or else we get the wrong results. Let me give you an example. From what we have seen in the previous section, 
\beq \int_{\Gamma_{d, a}} d \theta = a \sum_{l=1}^d e_l \eeq 
Therefore, 
\beq \int_{\Gamma_{d_1, a_1}} \bigg( d \theta_1 \cdot \int_{\Gamma_{d_2,a_2}} d \theta_2 \bigg) = \int_{\Gamma_{d_1,a_1}} \bigg(d \theta_1 \cdot \bigg(a_2 \sum_{l=1}^{d_2}e_l \bigg) \bigg) = \sum_{k=1}^{d_1} \int_0^{a_1} \bigg((dt \; e_k) \cdot \bigg( a_2 \sum_{k=1}^{d_2} e_k \bigg) \bigg) = \nonumber \eeq
\beq = a_2 \bigg( \int_0^{a_1} dt \bigg)  \sum_{l=1}^{d_2} e_k \cdot e_l  = a_1 a_2 \sum_{k=1}^{d_1} \sum_{l=1}^{d_2} e_k \cdot e_l = a_1a_2 \bigg( \sum_{k=l} e_k \cdot e_l + \sum_{k<l} e_k \cdot e_l + \sum_{k>l} e_k \cdot e_l \bigg)  = \nonumber \eeq 
\beq = a_1 a_2 \bigg( \sum_{k=1}^{\min (d_1, d_2)} 1 + \sum_{k<l} e_k \wedge e_l + \sum_{k>l} e_k \wedge e_l \bigg) = a_1 a_2 \bigg( \min (d_1, d_2) + 2 \sum_{k<l} e_k \wedge e_l \bigg) \label{Iterated} \eeq 
Now if we were to follow Eq \ref{PreferredChoice}, we would set $d_1=d_2=d$ and $a_1=a_2= (2/d)^{1/2}$, and obtain 
\beq \int_{\Gamma_{d, \sqrt{2/d}}} \bigg( d \theta_1 \cdot \int_{\Gamma_{d, \sqrt{2/d}}} d \theta_2 \bigg) = \frac{2}{d} \bigg(d + 2 \sum_{k<d}e_k \wedge e_l \bigg) = 2 + \frac{4}{d} \sum_{k<d}e_k \wedge e_l  \eeq
Therefore, 
\beq \lim^{max}_{d \rightarrow \infty}\int_{\Gamma_{d, \sqrt{2/d}}} \bigg( d \theta_1 \cdot \int_{\Gamma_{d, \sqrt{2/d}}} d \theta_2 \bigg) =2 \eeq 
which, of course, is bad since the choice of $a$ and $d$, given in Eq \ref{PreferredChoice}, was specifically designed to obtain the results expected from the conventional Grassmann integral; yet, in the present situation, we obtain $2$ despite the conventional answer being $0$. At the same time, it is still true, even in our framework, that 
\beq \lim^{max}_{d_1 \rightarrow \infty} \int_{\Gamma_{d_1, \sqrt{2/d_1}}} \bigg( d \theta_1 \cdot \bigg( \lim^{max}_{d_2 \rightarrow \infty} \int_{\Gamma_{d_2, \sqrt{2/d_2}}} d \theta_2 \bigg) \bigg) = \lim^{max}_{d_1 \rightarrow \infty} \int_{\Gamma_{d_1, \sqrt{2/d_1}}} (d \theta_1 \cdot 0) = 0 \eeq
The difference between those two cases is that, when we are taking two consequetive limits, we are implying that $1 \ll d_1 \ll d_2$, as opposed to the single limit that was implying $1 \ll d_1 =d_2$. If we go back to Eq \ref{Iterated} and plug in $a_1 = (2/d_1)^{1/2}$ and $a_2 = (2/d_2)^{1/2}$, we obtain 
\beq  \int_{\Gamma_{d_1, \sqrt{2/d_1}}} \bigg( d \theta_1 \cdot \int_{\Gamma_{d_2,\sqrt{2/d_2}}} d \theta_2 \bigg) =  \min \bigg(\sqrt{\frac{2}{d_1}} \sqrt{\frac{2}{d_2}} d_1, \sqrt{\frac{2}{d_1}} \sqrt{\frac{2}{d_2}} d_2 \bigg) + 2 \sqrt{\frac{2}{d_1}} \sqrt{\frac{2}{d_2}} \sum_{k<l} e_k \wedge e_l = \nonumber \eeq
\beq = 2 \sqrt{\min \bigg(\frac{d_1}{d_2}, \frac{d_2}{d_1}\bigg)} + \frac{4}{\sqrt{d_1d_2}} \sum_{k<l} e_k \wedge e_l \eeq 
That is why $d_1=d_2$ leads to the answer being $2$ whereas either $d_1 \ll d_2$ or $d_2 \ll d_1$ would lead to the answer being $0$ (due to "minimum" being taken). In other words, it doesn't matter in what order we take the limits, as long as the limits are consequetive as opposed to simultaneous; or, if we wanted to take the simultaneous limit, we could utilize the fact that 
\beq p < q \Longrightarrow d^p \ll d^q \; , \; q< p \Longrightarrow d^q \ll d^p \eeq
and write 
\beq \lim^{max}_{d \rightarrow \infty} \int_{\Gamma_{d^p, \sqrt{2/d^p}}} \bigg(d \theta_1 \cdot \int_{\Gamma_{d^q, \sqrt{2/d^q}}} d \theta_2 \bigg) = 2 \delta^p_q \eeq
which would give us the zero we want as long as $p \neq q$, regardless of which happens to be greater. On the other hand, if we take the integral of the form 
\beq \int (d \theta_1 \wedge d \theta_2) \cdot f (\theta_1, \theta_2) \eeq
then we would be able to obtain the correct answer independent of the contour:
\beq  \int (d \theta_1 \wedge d \theta_2) = 0 \eeq 
In contrast to Eq \ref{AlmostZero1Var}, the above is exact zero rather than approximate, and is independent of the contour. That is because $d \theta_1 \wedge d \theta_2$ is antisymmetric, and we couldn't have used antisymmetry in a single variable context. However, if we consider the integral of the form
\beq \int (d \theta_1 \wedge d \theta_2) \cdot \theta_1 \eeq
then we no longer have exact zero either, since $f (\theta_1, \theta_2)= \theta_1$ is not symmetric, in contrast to $f (\theta_1, \theta_2)=1$ which is. So, in the case of $f (\theta_1, \theta_2) = \theta_1$ we will again have to get a non-zero answer that approaches zero only in a limit. 

\section{Integrating $(d \theta \cdot e_k) \cdot \theta$ and $(d \theta \wedge e_k) \cdot\theta$}

In order to save ourselves some time, we would like to integrate $(d \theta \wedge e_k) \cdot \theta$ and $(d \theta \cdot e_k) \cdot \theta$ more or less at the same time, while keeping track of the differences between the two integrals. For that purpose, let us introduce the notation
\beq \delta^{\cdot}_{\cdot} = \delta^{\wedge}_{\wedge}=1 \; , \; \delta^{\cdot}_{\wedge} = \delta^{\wedge}_{\cdot} = 0 \eeq 
And, furthermore, let us define $*$ to be either $\cdot$ or $\wedge$:
\beq * \in (\cdot, \wedge) \eeq 
Thus, the integral we are interested in is 
\beq \int (d \theta * e_k) \cdot \theta \eeq
With this notation in mind, it is easy to see that 
\beq e_i * e_j = e_i \wedge e_j + \delta^i_j \delta^*_{\cdot} \eeq 
Therefore,
\beq \int_{\Gamma_{d,a}} (d \theta * e_k) \cdot \theta = \sum_{l=1}^d \int_0^a \bigg[ ((dt \; e_l)* e_k) \cdot \bigg(a \sum_{j=1}^{l-1} e_j + e_l t \bigg) \bigg] =  \nonumber \eeq
\beq = \sum_{l=1}^d \int_0^a \bigg[ dt \; (e_l \wedge e_k + \delta^k_l \delta^*_{\cdot}) \cdot \bigg(a \sum_{j=1}^{l-1} e_j + e_l t \bigg) \bigg] =  \nonumber \eeq
\beq = \sum_{l=1}^d \bigg[  \; (e_l \wedge e_k + \delta^k_l \delta^*_{\cdot}) \cdot \bigg(a \sum_{j=1}^{l-1} e_j \int_0^a dt + e_l \int_0^a t dt \bigg) \bigg] = \nonumber \eeq 
\beq = \sum_{l=1}^d \bigg[ (e_l \wedge e_k + \delta^k_l \delta^*_{\cdot}) \cdot \bigg(a^2 \sum_{j=1}^{l-1} e_j + \frac{a^2}{2} e_l \bigg) \bigg] = \nonumber \eeq 
\beq = a^2 \sum_{1 \leq j<l \leq d}( e_l \wedge e_k) \cdot e_j + \frac{a^2}{2} \sum_{l=1}^d (e_l \wedge e_k) \cdot e_l +  a^2 \delta^*_{\cdot} \sum_{1 \leq j <l \leq d} \delta^k_l e_j + \frac{a^2}{2} \delta^*_{\cdot} \sum_{l=1}^d \delta^k_l e_l \label{FourTerms}\eeq 
Let us look at the first sum. From the condition under the sum, we know that $j \neq l$. So the only question is whether or not either $j$ or $l$ is equal to $k$:
\beq k=j\neq l \Longrightarrow (e_l \wedge e_k) \cdot e_j =^{j=k} (e_l \wedge e_k) \cdot e_k = ^{l \neq k} (e_l \cdot e_k) \cdot e_k = e_l \cdot (e_k \cdot e_k) = e_l \cdot 1 = e_l \eeq 
\beq j\neq l=k \Longrightarrow (e_l \wedge e_k) \cdot e_j =^{l=k} (e_k \wedge e_k) \cdot e_j = 0 \cdot e_j = 0 \eeq 
\beq k \neq j \neq  l \neq k \Longrightarrow (e_l \wedge e_k) \cdot e_j = e_l \wedge e_k \wedge e_j \eeq 
The above can be summarized as 
\beq j \neq  l \Longrightarrow (e_l \wedge e_k) \cdot e_j = e_l \wedge e_k \wedge e_j  + e_l \delta^k_j \eeq
and, therefore, 
\beq \sum_{1 \leq j < l \leq d} (e_l \wedge e_k) \cdot e_j = \sum_{1 \leq j < l \leq d} e_l \wedge e_k \wedge e_j + \sum_{1 \leq j < l \leq d} e_l \delta^k_j = \sum_{1 \leq j <l \leq d} e_l \wedge e_k \wedge e_j + \sum_{l=k+1}^d e_l  \label{FirstTerm} \eeq
This equation covers both $k \leq d$ as well as $k>d$ if we define 
\beq a>b \Longrightarrow \sum_{l=a}^b (\cdots) = 0 \eeq 
Let us now look at the second term of Eq \ref{FourTerms}. We will compute the equation under the sum by cases:
\beq k = l \Longrightarrow (e_l \wedge e_k) \cdot e_l = (e_k \wedge e_k) \cdot e_k = 0 \cdot e_k = 0 \eeq 
\beq k \neq l \Longrightarrow (e_l \wedge e_k) \cdot e_l = - (e_k \wedge e_l) \cdot e_l =- (e_k \cdot e_l) \cdot e_l = - e_k \cdot (e_l \cdot e_l) = - e_k \cdot 1 = - e_k \eeq 
Thus, we can compute second term by cases as follows: 
\beq k \leq d \Longrightarrow \sum_{l=1}^d (e_l \wedge e_k) \cdot e_l = \sum_{l=1}^{k-1} (e_l \wedge e_k) \cdot e_l + (e_k \wedge e_k) \cdot e_k + \sum_{l=k+1}^d (e_l \wedge e_k) \cdot e_l = \nonumber \eeq 
\beq = \sum_{l=1}^{k-1} (-e_k) + \sum_{l=k+1}^d (-e_k) = - (d-1) e_k \eeq
\beq k> d \Longrightarrow  \sum_{l=1}^d (e_l \wedge e_k) \cdot e_l  = \sum_{l=1}^d (-e_k) = -d e_k \eeq 
If we now define the \emph{truth value} of a statement as 
\beq T (True) = 1 \; , \; T (False) =0 \eeq
the above two results generalize as 
\beq \sum_{l=1}^d (e_l \wedge e_k) \cdot e_l = - (d- T(k \leq d)) e_k \eeq 
Let us now look at the third term of Eq \ref{FourTerms}. Once again, we do that by cases:
\beq 1 \leq k \leq d \Longrightarrow \sum_{1 \leq j < l \leq d} \delta^k_l e_j = \sum_{1 \leq j < k} e_j = \sum_{j=1}^{k-1} e_j \eeq
\beq k > d \Longrightarrow \sum_{1 \leq j < l \leq d} \delta^k_l e_j = 0 \eeq 
Therefore, 
\beq \sum_{1 \leq j < l \leq d} \delta^k_l e_j = T (1 \leq k \leq d) \sum_{j=1}^{k-1} e_j \label{kljTruth} \eeq 
Finally, lets compute the last term: 
\beq 1 \leq k \leq d \Longrightarrow \sum_{l=1}^d \delta^k_l e_l = e_k \eeq
\beq k > d \Longrightarrow \sum_{l=1}^d \delta^k_l e_l = 0 \eeq
and, therefore, 
\beq \sum_{l=1}^d \delta^k_l e_l = e_k T (k \leq d) \eeq 
Thus, Eq \ref{FourTerms} becomes 
\beq \int_{\Gamma_{d,a}} (d \theta *e_k) \cdot \theta = a^2 \sum_{1 \leq j < l \leq d} e_l \wedge e_k \wedge e_j + a^2 \sum_{l=k+1}^d e_l  - \frac{a^2}{2} (d-T(k \leq d)) e_k + \nonumber \eeq
\beq + a^2 \delta^*_{\cdot} T (1 \leq k \leq d) \sum_{j=1}^{k-1} e_j + \frac{a^2}{2} \delta^*_{\cdot} e_k T (k \leq d) \eeq 
By noticing that 
\beq \sum_{l=k+1}^d e_l = T (1 \leq k \leq d) \sum_{l=k+1}^d e_l \eeq 
we can recombine the above terms to get 
\beq \int_{\Gamma_{d,a}} (d \theta * e_k) \cdot \theta = a^2 \sum_{1 \leq j < l \leq d} e_l \wedge e_k \wedge e_j + \nonumber \eeq
 \beq + a^2 T (1 \leq k \leq d) \bigg(\sum_{j=1}^{k-1} \delta^*_{\cdot} e_j + \sum_{l=k+1}^d e_l \bigg) - \frac{a^2}{2} (d- (1+ \delta^*_{\cdot}) T (k \leq d)) e_k \eeq 
Now, if we set 
\beq a = \sqrt{\frac{2}{d}} \eeq
then, in the limit of $d \rightarrow \infty$, all of the $a^2$ terms will go to zero with an exception of $a^2 d$ term. Therefore, we obtain
\beq \lim^{max}_{d \rightarrow \infty} \int_{\Gamma_{d, \sqrt{2/d}}} (d \theta * e_k) \cdot \theta =  - e_k \eeq 
where we have dropped $T (k \leq d)$ because, if $k=const$ then 
\beq \lim_{d \rightarrow \infty} T (k \leq d) =1 \label{LimitOfTruth} \eeq
Finally,  for any Grassmann number 
\beq \eta = \sum \eta_k e_k \eeq
we have 
\beq \lim^{max}_{d \rightarrow \infty} \int_{\Gamma_{d, \sqrt{2/d}}} (d \theta * \eta) \cdot \theta = - \eta \eeq 

\section{Integrating $(e_k*d\theta) \cdot \theta$} 

The integration of $(e_k * d \theta) \cdot \theta$ is very similar to the one of $(d \theta * e_k) \cdot \theta$ yet this won't allow us to skip the calculation altogether since there is some set of rather trivial differences that we have to keep track of. What we \emph{can} do, however, is simply compute the sum of the two results which would allow us to simply subtract the result of the previous section from that sum. We note that
\beq d \theta = e_j d t \Longrightarrow (e_k * d \theta) \cdot \theta + (d \theta * e_k) \cdot \theta = (e_k * e_j + e_j *e_k) \cdot \theta d t = 2 \delta^*_{\cdot} \delta^k_j  \cdot \theta dt = 2 \delta^*_{\cdot} \delta^k_j \theta d t \label{DiffToSimple} \eeq
where in the last step we used the fact that 
\beq c \in \mathbb{C} \Longrightarrow c \cdot \theta = c \theta \eeq 
Now, the fact that $d \theta = e_j dt$ implies that we know that we are on $j$-th edge of the contour and, therefore, 
\beq \theta = a \sum_{i=1}^{j-1} e_i + t e_j \eeq
which means that Eq \ref{DiffToSimple} can be rewritten as 
\beq (e_k * d \theta) \cdot \theta + (d \theta * e_k) \cdot \theta = 2 \delta^*_{\cdot} \delta^k_j d t \bigg(a \sum_{i=1}^{j-1} e_i + t e_j \bigg) \eeq 
Therefore, integrating the above expression gives us
\beq \int_{\Gamma_{d,a}} (e_k * d \theta) \cdot \theta + \int_{\Gamma_{d,a}} (d \theta * e_k) \cdot \theta = \sum_{j=1}^d \bigg( 2 \delta^*_{\cdot} \delta^k_j  \bigg(a \sum_{i=1}^{j-1} e_i \int_0^a dt +  e_j  \int_0^a t dt \bigg) \bigg) = \nonumber \eeq
\beq = \sum_{j=1}^d \bigg( 2 \delta^*_{\cdot} \delta^k_j \bigg( a \sum_{k=1}^{j-1} e_i a + e_j \frac{a^2}{2} \bigg) \bigg) =  2a^2 \delta^*_{\cdot} \bigg( \sum_{j=1}^d \delta^k_j \bigg)  \bigg( \sum_{i=1}^{j-1} e_i + \frac{e_j}{2} \bigg) = \nonumber \eeq
\beq = 2a^2 \delta^*_{\cdot}T (1 \leq k \leq d)  \bigg( \sum_{i=1}^{j-1} e_i + \frac{e_j}{2} \bigg)   \eeq
In the previous section we have obtained that 
\beq \int_{\Gamma_{d,a}} (d \theta * e_k) \cdot \theta = a^2 \sum_{1 \leq j < l \leq d} e_l \wedge e_k \wedge e_j + \nonumber \eeq
 \beq + a^2 T (1 \leq k \leq d) \bigg(\sum_{j=1}^{k-1} \delta^*_{\cdot} e_j + \sum_{l=k+1}^d e_l \bigg) - \frac{a^2}{2} (d- (1+ \delta^*_{\cdot}) T (k \leq d)) e_k \eeq 
and, therefore, we conclude
\beq \int_{\Gamma_{d,a}} (e_k * d \theta) \cdot \theta =  2 a^2 \delta^*_{\cdot} T(1 \leq k \leq d) \bigg(\sum_{l=1}^{k-1} e_l + \frac{e_k}{2} \bigg) - a^2 \sum_{1 \leq j < l \leq d} e_l \wedge e_k \wedge e_j + \nonumber \eeq
 \beq - a^2 T (1 \leq k \leq d) \bigg(\sum_{j=1}^{k-1} \delta^*_{\cdot} e_j + \sum_{l=k+1}^d e_l \bigg) + \frac{a^2}{2} (d- (1+ \delta^*_{\cdot}) T (k \leq d)) e_k \eeq 
If we now set 
\beq a = \sqrt{\frac{2}{d}} \eeq
then, due to the fact that all terms have $a^2$ factor, the only term that does \emph{not} go to zero is the one with $d$ in the numerator, that would cancel the $d$ in the denominator coming from $a^2$. By inspection of the above equation, we see that there is only one such term. Thus, we obtain
\beq \lim^{max}_{d \rightarrow \infty} \int_{\Gamma_{d, \sqrt{2/d}}} (e_k * d \theta ) \cdot \theta =  e_k \eeq
where we have dropped $T (k \leq d)$ due to Eq \ref{LimitOfTruth}.  

\section{Integrating $d \theta \cdot (\theta \wedge e_k)$ and $d \theta \cdot (\theta \wedge \eta)$} 

Let us now try to multiply $e_k$ by the finite part rather than by the differential. Since we know that the sign between the differential and the finite part is always a dot-product, while the sign within the finite part is always a wedge, there is only one way of doing it: namely, $d \theta \cdot (\theta \wedge e_k)$. Let us now go ahead and evaluate it:
\beq \int_{\Gamma_{d,a}} d \theta \cdot (\theta \wedge e_k) = \sum_{l=1}^d \int_0^a \bigg[(d t \; e_l) \cdot \bigg( \bigg(a \sum_{j=1}^{l-1} e_j + e_l t \bigg) \wedge e_k \bigg) \bigg] = \nonumber \eeq 
\beq = \sum_{l=1}^d  \bigg[ e_l \cdot \bigg( \bigg(a \sum_{j=1}^{l-1} e_j \int_0^a dt + e_l \int_0^a tdt \bigg) \wedge e_k \bigg) \bigg] = \nonumber \eeq 
\beq = \sum_{l=1}^d  \bigg[ e_l \cdot \bigg( \bigg(a^2  \sum_{j=1}^{l-1}  e_j  + e_l \frac{a^2}{2} \bigg) \wedge e_k \bigg) \bigg] = \nonumber \eeq 
\beq = a^2 \sum_{l=1}^d \sum_{j=1}^{l-1} e_l \cdot (e_j \wedge e_k) + \frac{a^2}{2} \sum_{l=1}^d e_l \cdot (e_l \wedge e_k) = \eeq 
\beq = a^2 \sum_{1 \leq j <l \leq d} e_l \cdot (e_j \wedge e_k) + \frac{a^2}{2} \sum_{l=1}^d e_l \cdot (e_l \wedge e_k) \label{TwoSums2} \eeq 
Let us look at the first sum. Since the condition under the sum implies $j \neq l$, the only question we have is whether $k = j$, or $k =l$, or neither. Let us look at all three cases:
\beq k=j \neq l \Longrightarrow e_l \cdot (e_j \wedge e_k) =^{k=j} e_l \cdot (e_k \wedge e_k) = e_l \cdot 0 =0 \eeq 
\beq j \neq l =k \Longrightarrow e_l \cdot (e_j \wedge e_k) = - e_l \cdot (e_k \wedge e_j) =^{k=l} -e_l \cdot (e_l \wedge e_j) =^{j \neq l} -e_l \cdot (e_l \cdot e_j) = \nonumber \eeq
 \beq =  - (e_l \cdot e_l) \cdot e_j = -1 \cdot e_j = - e_j \eeq 
\beq k \neq j \neq l \neq k \Longrightarrow e_l \cdot (e_j \wedge e_k) = e_l \wedge e_j \wedge e_k \eeq 
The above three equations generalize to 
\beq j \neq l \Longrightarrow e_l \cdot (e_j \wedge e_k) = e_l \wedge e_j \wedge e_k - e_j \delta^l_k \eeq 
Therefore, 
\beq \sum_{1 \leq j < l \leq d} e_l \cdot (e_j \wedge e_k) = \sum_{1 \leq j < l \leq d} (e_l \wedge e_j \wedge e_k - e_j \delta^l_k)  \eeq 
Now Eq \ref{kljTruth} tells us that 
\beq \sum_{1 \leq j < l \leq d} \delta^k_l e_j = T (1 \leq k \leq d) \sum_{j=1}^{k-1} e_j  \eeq 
Thus, we obtain 
\beq \sum_{1 \leq j < l \leq d} e_l \cdot (e_j \wedge e_k) = \sum_{1 \leq j < l \leq d} e_l \wedge e_j \wedge e_k - T (1 \leq k \leq d) \sum_{j=1}^{k-1} e_j \label{FirstTermTwoSums2} \eeq 
Let us now compute the second sum in Eq \ref{TwoSums2}. First we note that 
\beq l \neq k \Longrightarrow e_l \cdot (e_l \wedge e_k) = e_l \cdot (e_l \cdot e_k) = (e_l \cdot e_l) \cdot e_k = 1 \cdot e_k = e_k \eeq 
\beq l = k \Longrightarrow e_l \cdot (e_l \wedge e_k) = e_k \cdot (e_k \wedge e_k) = e_k \cdot 0 = 0 \eeq 
We then separate the cases of $k \leq d$ and $k >d$: 
\beq 1 \leq k \leq d \Longrightarrow \sum_{l=1}^d e_l \cdot (e_l \wedge e_k) = \sum_{l=1}^{k-1} e_l \cdot (e_l \wedge e_k) + 0 +  \sum_{l=k+1}^d e_l \cdot (e_l \wedge e_k)  = \nonumber \eeq
\beq = \sum_{l=1}^{k-1} e_k + \sum_{l=k+1}^d e_k =  e_k (d-1) \eeq 
\beq k > d \Longrightarrow \sum_{l=1}^d e_l \cdot (e_l \wedge e_k) = \sum_{l=1}^d e_k = e_k d \eeq 
Therefore, we can summarize it as 
\beq \sum_{l=1}^d e_l \cdot (e_l \wedge e_k) = e_k (d - T (1 \leq k \leq d)) \label{SecondTermTwoSums2} \eeq
Therefore, if we plug in \ref{FirstTermTwoSums2} and \ref{SecondTermTwoSums2} into Eq \ref{TwoSums2} we obtain
\beq \int_{\Gamma_{d,a}} d \theta \cdot (\theta \wedge e_k) = a^2 \sum_{1 \leq j < l \leq d} e_l \wedge e_j \wedge e_k - a^2 T (1 \leq k \leq d) \sum_{j=1}^{k-1} e_j + \nonumber \eeq
\beq + \frac{a^2d}{2} e_k - \frac{a^2}{2} e_k T (1 \leq k \leq d) \eeq
Finally, if we set 
\beq a = \sqrt{\frac{2}{d}} \eeq
then, in the limit of $d \rightarrow \infty$, all of the $a^2$ terms will be going to zero with an exception of the $a^2d$ term which will stay finite. As a result, we obtain
\beq \lim^{max}_{d \rightarrow \infty} \int_{\Gamma_{d, \sqrt{2/d}}} d \theta \cdot (\theta \wedge e_k) = e_k \eeq
Therefore, for a Grassmann number 
\beq \eta = \sum \eta_k e_k \eeq
we obtain 
\beq \lim^{max}_{d \rightarrow \infty} \int_{\Gamma_{d, \sqrt{2/d}}} d \theta \cdot (\theta \wedge \eta) = \eta \eeq 

\section{Integrating $(d \theta * e_i) \cdot (\theta \wedge e_j)$ where $i \neq j$}

In the previous two sections we tried $(d \theta * e_k) \cdot \theta$ and $d \theta \cdot  (\theta \wedge e_k)$. It is now time to try $(d \theta * e_i) \cdot (\theta \wedge e_j)$. In this section we will deal with the case of $i \neq j$, and we will leave $i =j$ for the next section.  Let us now go ahead and try to compute it. 
\beq \int_{\Gamma_{d,a}} (d \theta * e_i) \cdot (\theta \wedge e_j) = \sum_{k=1}^d \int_0^a \bigg[((d t \; e_k)*e_i) \cdot \bigg( \bigg(a \sum_{l=1}^{k-1} e_l + e_k t \bigg) \wedge e_j \bigg) \bigg] = \nonumber \eeq 
\beq = \sum_{k=1}^d  \bigg[(e_k*e_i) \cdot \bigg( \bigg(a \sum_{l=1}^{k-1} e_l \int_0^a dt + e_k \int_0^a tdt \bigg) \wedge e_j \bigg) \bigg] = \nonumber \eeq 
\beq = \sum_{k=1}^d  \bigg[(e_k*e_i) \cdot \bigg( \bigg(a^2 \sum_{l=1}^{k-1} e_l  + \frac{a^2}{2} e_k  \bigg) \wedge e_j \bigg) \bigg] = \nonumber \eeq 
\beq = a^2 \sum_{1 \leq l < k \leq d} (e_k * e_i) \cdot (e_l \wedge e_j) + \frac{a^2}{2} \sum_{k=1}^d (e_k * e_i) \cdot (e_k \wedge e_j) \label{ComplicatedTwoTerms} \eeq
Let us look at the first term. We know that $i \neq j$ from the title of this subsection, and we also know that $l \neq k$ from the condition under the sum. Finally we know that $j \neq l$ since we have the $e_l \wedge e_j$ factor. We summarize what we have just said as follows:
\beq {\rm First \; Term \; Of \; Eq \; \ref{ComplicatedTwoTerms}} \Longrightarrow i \neq j \neq l \neq k \label{FirstTermChain0} \eeq
So the three questions we have to ask is whether or not $k$ equals to $i$, whether or not $k$ equals to $j$, and whether or not $l$ equals to $i$. In other words, 
\beq l \; ? \; i \neq j \neq l \neq k \; ? \;  i \neq j \; ? \; k \label{FirstTermChain} \eeq 
and each of those three question marks needs to be replaced with either $=$ or $\neq$. Now those three replacements are not entirely independent of each other:
\beq (Eq \; \ref{FirstTermChain0}),  (i=l) \Longrightarrow^{l \neq k} i \neq k \label{Constraint1} \eeq
\beq (Eq \; \ref{FirstTermChain0}),  (j=k) \Longrightarrow^{i \neq j}  i \neq k \label{Constraint3} \eeq
Let us now count the number of options the above constraints rule out:

1. The constraint \ref{Constraint1} rules out $i=k=l$. However, if we were to have $i=k=l$ we could have either have $j=k$ or $j \neq k$. In other words it would consist of two options and we are ruling out BOTH of those two options. 

2. The constraint \ref{Constraint3} rules out an option $i=j=k$. In this case, again, there are two options: either $i=l$ or $i \neq l$. However the option $i=l$ will match the option $j=k$ from part 1. So we don't have to rule out the same option twice. Therefore, we are only ruling out ONE option: namely, the $j \neq k$ one. 

Therefore, the number of options left is 
\beq 2^3 - 2 -1=5 \eeq
and the list of those options is the following: 
\beq l = i \neq j \neq l \neq k \neq  i \neq j = k \label{chain1} \eeq 
\beq l = i \neq j \neq l \neq k \neq  i \neq j \neq k \label{chain2} \eeq 
\beq l \neq  i \neq j \neq l \neq k = i \neq j \neq  k \label{chain3} \eeq 
\beq l \neq  i \neq j \neq l \neq k \neq  i \neq j = k \label{chain4} \eeq 
\beq l \neq i \neq j \neq l \neq k \neq i \neq j \neq k \label{chain5} \eeq 
Let us now compute $(e_k*e_i) \cdot (e_l \wedge e_j)$ for each of those $5$ cases:
\beq l = i \neq j \neq l \neq k \neq  i \neq j = k \label{chain1} \Longrightarrow (e_k*e_i) \cdot (e_l \wedge e_j) = (e_j*e_i) \cdot (e_i \wedge e_j) =^{i \neq j} \nonumber \eeq
\beq =^{i \neq j} (e_j \cdot e_i) \cdot (e_i \cdot e_j)  = e_j \cdot (e_i \cdot e_i) \cdot e_j = e_j \cdot 1 \cdot e_j = e_j \cdot e_j = 1 \label{P}\eeq 
\beq l = i \neq j \neq l \neq k \neq  i \neq j \neq k \Longrightarrow (e_k*e_i) \cdot (e_l \wedge e_j) = (e_k*e_i) \cdot (e_i \wedge e_j) =^{k \neq i \neq j} \nonumber \eeq
\beq =^{k \neq i \neq j}  (e_k \cdot e_i) \cdot (e_i \cdot e_j) = e_k \cdot (e_i \cdot e_i) \cdot e_j = e_k \cdot 1 \cdot e_j = e_k \cdot e_j =^{k \neq j} e_k \wedge e_j \label{Q} \eeq 
\beq l \neq  i \neq j \neq l \neq k = i \neq j \neq  k \Longrightarrow (e_k*e_i) \cdot (e_l \wedge e_j) = (e_i*e_i) \cdot (e_l \wedge e_j) = \nonumber \eeq 
\beq = \delta^*_{\cdot} \cdot (e_l \wedge e_j) =  \delta^*_{\cdot} e_l \wedge e_j \label{R} \eeq 
\beq l \neq  i \neq j \neq l \neq k \neq  i \neq j = k \Longrightarrow (e_k*e_i) \cdot (e_l \wedge e_j) = (e_j*e_i) \cdot (e_l \wedge e_j) = \nonumber \eeq 
\beq =^{i \neq j} (e_j \wedge e_i) \cdot (e_l \wedge e_j) = (e_i \wedge e_j) \cdot (e_j \wedge e_l) = ^{i \neq j \neq l} (e_i \cdot e_j) \cdot (e_j \cdot e_l) = \nonumber \eeq
\beq = e_i \cdot (e_j \cdot e_j) \cdot e_l = e_i \cdot 1 \cdot e_l = e_i \cdot e_l =^{l \neq i} e_i \wedge e_l \label{S} \eeq
\beq l \neq i \neq j \neq l \neq k \neq i \neq j \neq k \Longrightarrow (e_k*e_i) \cdot (e_l \wedge e_j) = e_k \wedge e_i \wedge e_l \wedge e_j \label{T} \eeq 
We will now sum over all of the above combinations:
\beq i \neq j \Longrightarrow \sum_{1 \leq l < k \leq d} (e_k *e_i) \cdot (e_l \wedge e_j)= T (1 \leq i <j \leq d)+ \nonumber\eeq
\beq + T (1 \leq i \leq d) \sum_{k=i+1}^d e_k \wedge e_j + \delta^*_{\cdot} T (1 \leq i \leq d) \sum_{l=1}^{i-1} e_l \wedge e_j + \nonumber \eeq
 \beq + T (1 \leq j \leq d) \sum_{l=1}^{j-1} e_i \wedge e_l  +    \sum_{1 \leq l < k \leq d}  e_k \wedge e_i \wedge e_l \wedge e_j \label{Part1OfLongestSum0} \eeq
where, in each case, we have used the fact that either $i$ and/or $j$ is equal to either $k$ and/or $l$ in order to read off the conditions for $i$ and $j$ (under $T$-functions) from our knowledge that $1 \leq l < k \leq j$. The fact that we have $k < l$ instead of $k \neq l$ is the reason why in the 2-nd, 3-rd and 4-th terms we have terminated the sums instead of simply skipping over one term. 

Intuitively, we can make sense of this in terms of "contractions": whenever two indexes happened to be equal, they get contracted with each other and both disappear. Since the condition under the sum implies $k \neq l$, we know that $k$ can't be contracted with $l$. Instead, we can contract $k$ with either $i$ or $j$. Contracting $k$ with $i$ produces the third term, and contracting $k$ with $j$ produces the first and fourth term  (in the first term, in addition to that contraction, $i$ is also gets contracted with $l$, while in the fourth term $i$ and $l$ remain un-contracted). Thus, we have exhausted all of the ways of contracting $k$. Now, if we don't contract $k$, we can either contract $i$ or not. Now, we know that $i$ can't be contracted with $j$ since we have $i \neq j$ in the title of this section. Therefore, if we wish to contract $i$, the only way of doing so is to contract it with $l$. Now, if we contract $i$ with $l$, then the only way of contracting $k$ would be with $j$, which would bring us back to the first term (which we have already covered earlier). The only other option in case of $i$ being contracted with $l$ is not to contract $k$ at all, in which case we would get the second term. Thus, we have covered all of the ways of contracting \emph{either} $k$ \emph{or} $i$ \emph{or both}. Finally, if we \emph{neither} contract $k$ \emph{nor} $i$, then the only way to contract $l$ and $j$ is to contract them with each other; but we can't do that since we know that $l \neq j$ due to the wedge product. Thus, the only option is the last term. In other words, every single way of contracting the indexes given the above conditions would return to us one of the terms on the right hand side, which is why we don't have any other terms. 

What we computed so far is only the first term on the right hand side of Eq \ref{ComplicatedTwoTerms}. Let us now compute the second term, 
\beq \sum_{k=1}^d (e_k * e_i) \cdot (e_k \wedge e_j)  \nonumber \eeq 
Now we have considerably fewer cases to work out. From the wedge product $e_k \wedge e_j$ we know that $k \neq j$ and also from the title of this section we know that $i \neq j$. Therefore, 
 \beq {\rm Second \; Term \; Of \; Eq \; \ref{ComplicatedTwoTerms}} \Longrightarrow i \neq j \neq k \eeq 
Thus, the only question is whether $i =k$ or $i \neq k$, which leaves us at only two cases, $k= i \neq j \neq k$ and $k \neq i \neq j \neq k$. 
\beq k= i \neq j \neq k \Longrightarrow (e_k *e_i) \cdot (e_k \wedge e_j) = (e_i * e_i) \cdot (e_i \wedge e_j) = \delta^*_{\cdot} \cdot e_i \wedge e_j = \delta^*_{\cdot} e_i \wedge e_j \eeq
\beq k \neq i \neq j \neq k \Longrightarrow (e_k*e_i) \cdot (e_k \wedge e_j) =^{k \neq i} (e_k \wedge e_i) \cdot (e_k \wedge e_j) = - (e_i \wedge e_k) \cdot (e_k \wedge e_j) \nonumber \eeq
\beq =^{i \neq k \neq j} - (e_i \cdot e_k) \cdot (e_k \cdot e_j) = - e_i \cdot (e_k \cdot e_k) \cdot e_j  = - e_i \cdot 1 \cdot e_j = - e_i \cdot e_j =^{i \neq j} - e_i \wedge e_j \eeq 
Now, in the previous case we were contracting with either $i$ or $j$ and there was only one copy of each of them. On the other hand, right now we are contracting $k$ with itself, and there are multiple copies of $k$; so we have to count all of them and put it as a coefficient. In the case of $k = i \neq j \neq k$, there is only one option for $k$, namely $k=i$ if $1 \leq i \leq d$ and zero options if $i>d$. Thus, the coefficient is $T (1 \leq i \leq d)$. On the other hand, in the case $k \neq i \leq j \leq k$ the number of ways of picking $k$ is $d - T (1 \leq i \leq d) - T (1 \leq j \leq d)$, and that would be the coefficient. Therefore, we obtain 
\beq i \neq j \Longrightarrow \sum_{k=1}^d (e_k * e_i) \cdot (e_k \wedge e_j) = \delta^*_{\cdot} e_i \wedge e_j T (1 \leq i \leq d) - \nonumber \eeq
\beq -  e_i \wedge e_j (d - T (1 \leq i \leq d) - T (1 \leq j \leq d)) \eeq 
Since both terms have $e_i \wedge e_j$, we can combine them and obtain
\beq i \neq j \Longrightarrow \sum_{k=1}^d (e_k * e_i) \cdot (e_k \wedge e_j) = - e_i \wedge e_j (d - (1- \delta^*_{\cdot}) T (1 \leq i \leq d) - T (1 \leq j \leq d)) \eeq 
By noticing that 
\beq 1 - \delta^*_{\cdot} = \delta^*_{\wedge} \eeq
we can further rewrite it as 
\beq i \neq j \Longrightarrow \sum_{k=1}^d (e_k * e_i) \cdot (e_k \wedge e_j) = - e_i \wedge e_j (d - \delta^*_{\wedge} T (1 \leq i \leq d) - T (1 \leq j \leq d)) \eeq 
Therefore, Eq \ref{ComplicatedTwoTerms} becomes 
\beq i \neq j \Longrightarrow \int_{\Gamma_{d,a}} (d \theta * e_i) \cdot (\theta \wedge e_j) =  a^2 \sum_{1 \leq l < k \leq d} (e_k *e_i) \cdot (e_l \wedge e_j)= a^2 T (1 \leq i <j \leq d)+ \nonumber\eeq
\beq + a^2 T (1 \leq i \leq d) \sum_{k=i+1}^d e_k \wedge e_j + a^2 \delta^*_{\cdot} T (1 \leq i \leq d) \sum_{l=1}^{i-1} e_l \wedge e_j + \nonumber \eeq
 \beq + a^2 T (1 \leq j \leq d) \sum_{l=1}^{j-1} e_i \wedge e_l  +    a^2 \sum_{1 \leq l < k \leq d}  e_k \wedge e_i \wedge e_l \wedge e_j + \nonumber \eeq
\beq - e_i \wedge e_j (a^2 d - a^2 \delta^*_{\wedge} T (1 \leq i \leq d) - a^2 T (1 \leq j \leq d)) \eeq 
Finally, if we set
\beq a = \sqrt{\frac{2}{d}} \eeq
then, in the limit of $d \rightarrow \infty$, all of the $a^2$-terms will disappear while the $a^2d$ will be replaced with $2$. As a result, we obtain
\beq i \neq j \Longrightarrow \lim^{max}_{d \rightarrow \infty} \int_{\Gamma_{d, \sqrt{2/d}}} (d \theta *e_j) \cdot (\theta \wedge e_i) = - e_i \wedge e_j \eeq 

\section{Integrating $(d \theta * e_i) \cdot (\theta \wedge e_j) = (d \theta * e_i) \cdot (\theta \wedge e_i)$, where $i =j$} 

Since in the previous section we were explicitly assuming $i \neq j$, we will now have to separately cover the $i=j$ case. Since Eq \ref{ComplicatedTwoTerms} was not based on that assumption, we can rewrite its result while substituting $i=j$:
\beq \int_{\Gamma_{d,a}} (d \theta * e_i) \cdot (\theta \wedge e_i)= a^2 \sum_{1 \leq l < k \leq d} (e_k * e_i) \cdot (e_l \wedge e_i) + \frac{a^2}{2} \sum_{k=1}^d (e_k * e_i) \cdot (e_k \wedge e_i)  \label{iISj} \eeq
Now, as far as the first term goes, we know that $l \neq k$ from the condition under the sum. Furthermore, we know that $l \neq i$ from $e_l \wedge e_i$. Thus, we know that $i \neq l \neq k$. The only question is whether $i=k$ or $i \neq k$. Thus, we have two cases: $i \neq l \neq k =i$ and $i \neq l \neq k \neq i$. 
\beq i \neq l \neq k = i \Longrightarrow (e_k *e_i) \cdot (e_l \wedge e_i) = (e_i * e_i) \cdot (e_l \wedge e_i) = \delta^*_{\cdot} \cdot (e_l \wedge e_i) = \delta^*_{\cdot} e_l \wedge e_i \eeq 
\beq i \neq l \neq k \neq i \Longrightarrow (e_k * e_i) \cdot (e_l \wedge e_i) =^{k \neq i} (e_k \wedge e_i) \cdot (e_l \wedge e_i) = - (e_k \wedge e_i) \cdot (e_i \wedge e_l) =^{k \neq i \neq l} \nonumber \eeq
\beq  =^{k \neq i \neq l}  - (e_k \cdot e_i) \cdot (e_i \cdot e_l) = - e_k \cdot (e_i \cdot e_i) \cdot e_l = - e_k \cdot 1 \cdot e_l = - e_k \cdot e_l =^{k \neq l} - e_k \wedge e_l \eeq 
Therefore, 
\beq \sum_{1 \leq l < k \leq d} (e_k * e_i) \cdot (e_l \wedge e_i) = \delta^*_{\cdot} \sum_{l=1}^d e_l \wedge e_i - \sum_{k,l \in \{1, \cdots, d\} \setminus \{i \}} e_k \wedge e_l T (l<k) \eeq
Let us now look at the second term of Eq \ref{iISj}. In this case $e_k \wedge e_i$ implies $k \neq i$ and, since no other letters are used, the latter is the only option. Thus, 
\beq k \neq i \Longrightarrow (e_k *e_i) \cdot (e_k \wedge e_i) =^{k \neq i} (e_k \wedge e_i) \cdot (e_k \wedge e_i) = - (e_k \wedge e_i) \cdot (e_i \wedge e_k) =^{k \neq i} \nonumber \eeq
\beq  =^{k \neq i}  - (e_k \cdot e_i) \cdot (e_i \cdot e_k) = - e_k \cdot (e_i \cdot e_i) \cdot e_k = - e_k \cdot 1 \cdot e_k = - e_k \cdot e_k = -1 \eeq 
Since there are $d-T (1 \leq i \leq d)$ copies of it, coming from the number of values of $k \neq i$, we have
\beq \sum_{k=1}^d (e_k *e_i) \cdot (e_k \wedge e_i) = - (d- T (1 \leq i \leq d)) \eeq 
Therefore, Eq \ref{iISj} becomes 
\beq \int_{\Gamma_{d,a}} (d \theta * e_i) \cdot (\theta \wedge e_i) = a^2 \delta^*_{\cdot} \sum_{l=1}^d e_l \wedge e_i - \nonumber \eeq
\beq - a^2 \sum_{k,l \in \{1, \cdots, d \} \setminus \{i \}} e_k \wedge e_l T (l<k)  - \frac{a^2 d - a^2 T(1 \leq i \leq d)}{2} \eeq 
Now, if we set
\beq a = \sqrt{\frac{2}{d}} \eeq
then in the limit of $d \rightarrow \infty$ all of the $a^2$ terms will disappear, except for $a^2d$, which will become $2$, thus we obtain
\beq \lim^{max}_{d \rightarrow \infty} \int_{\Gamma_{d, \sqrt{2/d}}} (d \theta * e_i) \cdot (\theta \wedge e_j) = -1 \eeq
The fact that this is $-1$ instead of $+1$ is related to the minus sign we will see in Eq \ref{WrongSingMarch} which, in turn, is related to the sign issue discussed in Section \ref{SignSection} 

\section{Integrating $(d \theta_1 * d \theta_2) \cdot \theta_1$}

Let us now integrate $(d \theta_1 * d \theta_2) \cdot \theta_1$. Unline the previous integrals, we now have two contours: $d \theta_1$ is integrated over $\Gamma_{d_1a_1}$ and $d \theta_2$ is integrated over $\Gamma_{d_2a_2}$. We then extract $e_k$ out of $d \theta_2$ (where $k$ depends on what part of the contour $\theta_2$ happens to be at) and then integrating $(d \theta_1 *e_k) \cdot \theta_1$. Therefore, 
\beq \int_{\theta_1 \in \Gamma_{d_1,a_1} ; \theta_2 \in \Gamma_{d_2,a_2}}  (d \theta_1 * d \theta_2) \cdot \theta_1 \eeq
We have found earlier that 
\beq \int_{\Gamma_{d,a}} (d \theta * e_k) \cdot \theta = a^2 \sum_{1 \leq j < l \leq d} e_l \wedge e_k \wedge e_j + \nonumber \eeq
 \beq + a^2 T (1 \leq k \leq d) \bigg(\sum_{j=1}^{k-1} \delta^*_{\cdot} e_j + \sum_{l=k+1}^d e_l \bigg) - \frac{a^2}{2} (d- (1+ \delta^*_{\cdot}) T (k \leq d)) e_k \eeq 
Therefore,
\beq \int_{\theta_1 \in \Gamma_{d_1,a_1} ; \theta_2 \in \Gamma_{d_2,a_2}}  (d \theta_1 * d \theta_2) \cdot \theta_1 = a_2 \sum_{k=1}^{d_2} \int_{\Gamma_{d_1a_1}}  (d \theta_1 * e_k) \cdot \theta_1 = \nonumber \eeq
\beq = a_2 \sum_{k=1}^{d_2} \bigg(a_1^2 \sum_{1 \leq j < l \leq d_1} e_l \wedge e_k \wedge e_j + \nonumber \eeq
 \beq + a_1^2 T (1 \leq k \leq d_1) \bigg(\sum_{j=1}^{k-1} \delta^*_{\cdot} e_j + \sum_{l=k+1}^{d_1} e_l \bigg) - \frac{a_1^2}{2} (d_1- (1+ \delta^*_{\cdot}) T (k \leq d_1)) e_k \bigg)  \label{Sunny} \eeq 
Let us compute the first term:
\beq a_2  \sum_{k=1}^{d_2} \bigg( a_1^2\sum_{1 \leq j <l \leq d_1} e_l \wedge e_k \wedge e_j \bigg) = a_2 a_1^2  \sum_{k=1}^{d_2}  \sum_{1 \leq j <l \leq d_1}  e_l \wedge e_k \wedge e_j \eeq
The rest of the terms have single $e$, so we have to pay attention to how many times each $e$ occurs. We can get rid of the factor $T (1 \leq k \leq d_1)$ on the second term by simply changing the condition under the sum from $1 \leq k \leq d_2$ to $1 \leq k \leq \min (d_1, d_2)$. Keeping this in mind, we can do the following calculation: 
\beq a_2 \sum_{k=1}^{d_2} \bigg(a_1^2 T(1 \leq k \leq d_1) \sum_{j=1}^{k-1} \delta^*_{\cdot} e_j \bigg) = a_2 a_1^2 \sum_{k=1}^{\min (d_1, d_2)} \sum_{j=1}^{k-1} \delta^*_{\cdot} e_j = \nonumber \eeq 
\beq = a_2 a_1^2 \delta^*_{\cdot} \sum_{1 \leq j <k \leq \min (d_1, d_2)}  e_j = a_2 a_1^2 \delta^*_{\cdot} \sum_{j=1}^{\min (d_1, d_2)} e_j (\min (d_1, d_2) - j)\eeq
Similarly, the third term of Eq \ref{Sunny} evaluates to 
\beq a_2 \sum_{k=1}^{d_2} \bigg(a_1^2 T(1 \leq k \leq d_1) \sum_{l=k+1}^{d_1} e_l \bigg) = a_2 a_1^2 \sum_{k=1}^{\min (d_1,d_2)} \sum_{l=k+1}^{d_1} e_l = \nonumber \eeq
\beq = a_2 a_1^2 \sum_{l=2}^{d_1} \sum_{k=1}^{\min (d_1, d_2, l-1)} e_l = a_2 a_1^2 \sum_{l=2}^{d_1} (e_l \; \min (d_1, d_2, l-1)) \eeq
The fourth term is 
\beq a_2 \sum_{k=1}^{d_2} \bigg(- \frac{a_1^2}{2} d_1 e_k \bigg) = - \frac{a_2a_1^2d_1}{2} \sum_{k=1}^d e_k \eeq 
and the fifth term is 
\beq a_2 \sum_{k=1}^{d_2} \bigg(- \frac{a_1^2}{2} (- (1 + \delta^*_{\cdot}) T (1 \leq d_1) e_k) \bigg) = \frac{a_2 a_1^2}{2} (1 + \delta^*_{\cdot}) \sum_{k=1}^{\min (d_1, d_2)} e_k \eeq 
Thus, putting all those terms together, Eq \ref{Sunny} evaluates to 
\beq \int_{\theta_1 \in \Gamma_{d_1,a_1} ; \theta_2 \in \Gamma_{d_2,a_2}}  (d \theta_1 * d \theta_2) \cdot \theta_1 = a_2 a_1^2  \sum_{k=1}^{d_2} \sum_{1 \leq j <l \leq d_1} e_l \wedge e_k \wedge e_j +a_2 a_1^2 \delta^*_{\cdot} \sum_{j=1}^{\min (d_1, d_2)} e_j (\min (d_1, d_2) - j) + \nonumber \eeq
\beq + a_2 a_1^2 \sum_{l=2}^{d_1} (e_l \; \min (d_1, d_2, l-1)) - \frac{a_2a_1^2d_1}{2} \sum_{k=1}^d e_k +\frac{a_2 a_1^2}{2} (1 + \delta^*_{\cdot}) \sum_{k=1}^{\min (d_1, d_2)} e_k \eeq
Now, if we set 
\beq a_1 = \sqrt{\frac{2}{d_1}} \;, \; a_2 = \sqrt{\frac{2}{d_2}} \eeq
then the 1-st, 2-nd, 3-rd and 5-th terms go trivially to zero. As far as the 4-th term, $a_1^2d_1$ becomes $2$, but then the extra factor of $a_2$ sends it to zero. Thus, the total sum is sent to zero as well: 
\beq \lim^{max}_{d_1 \rightarrow \infty, d_2 \rightarrow \infty} \int_{\theta_1 \in \Gamma_{d_1, \sqrt{2/d_1}}, \theta_2 \in \Gamma_{d_2, \sqrt{2/d_2}}}  (d \theta_1 * d \theta_2) \cdot \theta_1 = 0 \eeq 

\section{Integration $(d \theta_1 * d \theta_2) \cdot \theta_2$} 

The integration of $(d \theta_1 * d \theta_2) \cdot \theta_2$ is similar to the $(d \theta_1 * d \theta_2) \cdot \theta_1$, yet there are some trivial differences between the two expressions. In order not to have to repeat a very similar calculation, we will use the following trick. First, we compute the sum of the integrals of $(d \theta_1 * d \theta_2) \cdot \theta_2$ and $(d \theta_2 * d \theta_1) \cdot \theta_2$. Then we will re-label the indexes in the previous section to obtain the integral of $(d \theta_2 * d \theta_1) \cdot \theta_2$. Finally, by subtracting the latter from the sum of the two integrals, we will obtain the integral of $(d \theta_1 * d \theta_2) \cdot \theta_2$. 

Let us go ahead and compute the sum of the two integrals. Given the definition of contours $\Gamma_{d_1,a_1}$ and $\Gamma_{d_2, a_2}$, we know that $d \theta_1$ and $d \theta_2$ are either perpendicular or parallel to each other. If they happened to be perpendicular to each other, then $d \theta_1 * d \theta_2$ and $d \theta_2 * d \theta_1$ will be replaced with $d \theta_1 \wedge d \theta_2$ and $d \theta_2 \wedge d \theta_1$, which means that their sum will be zero (and $\theta_2$ will simply be factored out of the sum as a common factor). This means that the only terms that survive are the ones where $d \theta_1$ and $d \theta_2$ are parallel to each other. In order for them to be parallel to each other, they have to reside on the edge number $j$ of their respective contours, where $j$ is the same number, despite the fact that the contours are different. In order for $d \theta_1$ to reside on edge number $j$, we need $1 \leq j \leq d_1$ and in order for $d \theta_2$ to reside on the edge number $j$ we need $1 \leq j \leq d_2$. In order for those two conditions to simultaneously be true, we need 
\beq 1 \leq j \leq \min (d_1, d_2) \eeq
As long as $d \theta_1$ and $d \theta_2$ both reside on the edge $j$, we have 
\beq {\rm Same \; Edge} \; \Longrightarrow \; d \theta_1 * d \theta_2 = (e_j dt_1) * (e_j dt_2) = dt_1 dt_2 e_j *e_j = dt_1 dt_2 \delta^*_{\cdot} \eeq
and, therefore, we can evaluate the sum of the integrals as follows: 
\beq \int_{\theta_1 \in \Gamma_{d_1,a_1}, \theta_2 \in \Gamma_{d_2,a_2}} (d \theta_1 * d \theta_2) \cdot \theta_2 + \int_{\theta_1 \in \Gamma_{d_1,a_1}, \theta_2 \in \Gamma_{d_2,a_2}} (d \theta_2 * d \theta_1) \cdot \theta_2 = \nonumber \eeq 
\beq = \sum_{j=1}^{\min (d_1,d_2)} \int_0^{a_1} dt_1 \int_0^{a_2} dt_2 (e_j *e_j) \cdot \bigg(t_2 e_j + a_2 \sum_{i=1}^{j-1} e_i \bigg) = \nonumber \eeq 
\beq = \sum_{j=1}^{\min (d_1,d_2)} \int_0^{a_1} dt_1 \int_0^{a_2} dt_2 \delta^*_{\cdot} \cdot \bigg(t_2 e_j + a_2 \sum_{i=1}^{j-1} e_i \bigg) = \nonumber \eeq 
\beq = \sum_{j=1}^{\min (d_1,d_2)}  \delta^*_{\cdot} \bigg( \int_0^{a_1} dt_1 \bigg) \bigg(\bigg( \int_0^{a_2} t_2 dt_2 \bigg) e_j + a_2 \sum_{i=1}^{j-1} e_i \bigg(\int_0^{a_2} dt_2 \bigg) \bigg) = \nonumber \eeq 
\beq = \sum_{j=1}^{\min (d_1,d_2)} \delta^*_{\cdot} a_1 \bigg( \frac{a_2^2}{2} e_j + a_2 \sum_{i=1}^{j-1} e_i a_2 \bigg) = \delta^*_{\cdot} a_1 a_2^2 \bigg( \frac{1}{2} \sum_{j=1}^{\min (d_1, d_2)} e_j + \sum_{1 \leq i <j \leq \min (d_1,d_2)} e_i \bigg)  \label{SumForSimplicity} \eeq 
In the previous section, we have found that 
\beq \int_{\theta_1 \in \Gamma_{d_1,a_1} ; \theta_2 \in \Gamma_{d_2,a_2}}  (d \theta_1 * d \theta_2) \cdot \theta_1 = a_2 a_1^2  \sum_{k=1}^{d_2} \sum_{1 \leq j <l \leq d_1} e_l \wedge e_k \wedge e_j +a_2 a_1^2 \delta^*_{\cdot} \sum_{j=1}^{\min (d_1, d_2)} e_j (\min (d_1, d_2) - j) + \nonumber \eeq
\beq + a_2 a_1^2 \sum_{l=2}^{d_1} (e_l \; \min (d_1, d_2, l-1)) - \frac{a_2a_1^2d_1}{2} \sum_{k=1}^d e_k +\frac{a_2 a_1^2}{2} (1 + \delta^*_{\cdot}) \sum_{k=1}^{\min (d_1, d_2)} e_k \eeq
therefore, if we re-label the indexes, we obtain
\beq \int_{\theta_1 \in \Gamma_{d_1,a_1} ; \theta_2 \in \Gamma_{d_2,a_2}}  (d \theta_2 * d \theta_1) \cdot \theta_2 = a_1 a_2^2  \sum_{k=1}^{d_1} \sum_{1 \leq j <l \leq d_2} e_l \wedge e_k \wedge e_j +a_1 a_2^2 \delta^*_{\cdot} \sum_{j=1}^{\min (d_1, d_2)} e_j (\min (d_1, d_2) - j) + \nonumber \eeq
\beq + a_1 a_2^2 \sum_{l=2}^{d_1} (e_l \; \min (d_1, d_2, l-1)) - \frac{a_1a_2^2d_2}{2} \sum_{k=1}^d e_k +\frac{a_1 a_2^2}{2} (1 + \delta^*_{\cdot}) \sum_{k=1}^{\min (d_1, d_2)} e_k \eeq
which, in combination with Eq \ref{SumForSimplicity} produces
\beq \int_{\theta_1 \in \Gamma_{d_1,a_1} ; \theta_2 \in \Gamma_{d_2,a_2}} (d \theta_1 * d \theta_2) \cdot \theta_2   = \delta^*_{\cdot} a_1 a_2^2 \bigg( \frac{1}{2} \sum_{j=1}^{\min (d_1, d_2)} e_j + \sum_{1 \leq i <j \leq \min (d_1,d_2)} e_i \bigg)  -\eeq 
\beq - a_1 a_2^2  \sum_{k=1}^{d_1} \sum_{1 \leq j <l \leq d_2} e_l \wedge e_k \wedge e_j -a_1 a_2^2 \delta^*_{\cdot} \sum_{j=1}^{\min (d_1, d_2)} e_j (\min (d_1, d_2) - j) - \nonumber \eeq
\beq - a_1 a_2^2 \sum_{l=2}^{d_1} (e_l \; \min (d_1, d_2, l-1)) + \frac{a_1a_2^2d_2}{2} \sum_{k=1}^d e_k - \frac{a_1 a_2^2}{2} (1 + \delta^*_{\cdot}) \sum_{k=1}^{\min (d_1, d_2)} e_k \eeq
Now, if we set 
\beq a_1 = \sqrt{\frac{2}{d_1}} \; , \; a_2 = \sqrt{\frac{2}{d_2}} \eeq
then, by noting that every single term contains $a_1a_2^2$, we need some extra factors of $d_1$ and $d_2$ in the numerator in order to prevent any given term from going to zero as $d_1 \rightarrow \infty$ and $d_2 \rightarrow \infty$. The only term with $d$ in the numerator is the second before the end. But even then it doesn't have enough $d$-s: after all, $d_2$ neutralizes the effect of $a_2^2$ via $a_2^2d_2=2$ yet we don't have any $d$-s to neutralize the effect of $a_1$, so that we still have
\beq \frac{a_1a_2^2d_2}{2} = a_1 = \sqrt{2}{d_1} \rightarrow 0 \eeq
Therefore, we have 
\beq \lim^{max}_{d_1 \rightarrow \infty, d_2 \rightarrow \infty} \int_{\theta_1 \in \Gamma_{d_1,a_1} ; \theta_2 \in \Gamma_{d_2,a_2}} (d \theta_1 * d \theta_2) \cdot \theta_2 = 0 \eeq 

\section{Integrating $(d \theta_1 * d \theta_2) \cdot (\theta_1 \wedge \theta_2)$}

If we assume that $\theta_1 \in \Gamma_{d_1, a_1}$ and $\theta_2 \in \Gamma_{d_2, a_2}$ then it is easy to see that 
\beq d \theta_2 = e_i dt \eeq
\beq \theta_2 = e_i t + a_2 \sum_{j=1}^{i-1} e_j \eeq
and, therefore, we obtain 
\beq \int_{\theta_1 \in  \Gamma_{d_1, a_1}, \theta_2 \in \Gamma_{d_2, a_2}} (d \theta_1 * d \theta_2) \cdot (\theta_1 \wedge \theta_2)= \nonumber \eeq
\beq = \sum_{i=1}^{d_2}  \int_{\theta_1 \in  \Gamma_{d_1, a_1}} \int_0^{a_2} (d \theta_1 * (e_i dt)) \cdot \bigg(\theta_1 \wedge \bigg(e_i t + a_2 \sum_{j=1}^{i-1} e_j \bigg) \bigg)= \nonumber \eeq
\beq = \sum_{i=1}^{d_2}  \int_{\theta_1 \in  \Gamma_{d_1, a_1}} (d \theta_1 * e_i ) \cdot \bigg(\theta_1 \wedge \bigg(e_i \int_0^{a_2} t dt + a_2 \sum_{j=1}^{i-1} e_j \int_0^{a_2} dt \bigg) \bigg)= \nonumber \eeq
\beq = \sum_{i=1}^{d_2}  \int_{\theta_1 \in  \Gamma_{d_1, a_1}} (d \theta_1 * e_i ) \cdot \bigg(\theta_1 \wedge \bigg(e_i \frac{a_2^2}{2} + a_2^2 \sum_{j=1}^{i-1} e_j  \bigg) \bigg)= \nonumber \eeq
\beq =\frac{a_2^2}{2} \sum_{i=1}^{d_2} \int_{\theta_1 \in \Gamma_{d_1,a_1}} (d \theta_1 * e_i) \cdot (\theta_1 \wedge e_i)  + a_2^2 \sum_{1 \leq j <i \leq d_2}  \int_{\theta_1 \in \Gamma_{d_1,a_1}} (d \theta_1 *e_i) \cdot (\theta_1 \wedge e_j)  \eeq
Now, we know from previous results that 
\beq \int_{\Gamma_{d,a}} (d \theta * e_i) \cdot (\theta \wedge e_i) = a^2 \delta^*_{\cdot} \sum_{l=1}^d e_l \wedge e_i - \nonumber \eeq
\beq -  a^2 \sum_{k,l \in \{1, \cdots, d \} \setminus \{i \}} e_k \wedge e_l T (l<k) - \frac{a^2 d - a^2 T(1 \leq i \leq d)}{2} \eeq 
\beq i \neq j \Longrightarrow \int_{\Gamma_{d,a}} (d \theta * e_i) \cdot (\theta \wedge e_j) =  a^2 T (1 \leq i <j \leq d)+ \nonumber\eeq
\beq + a^2 T (1 \leq i \leq d) \sum_{k=i+1}^d e_k \wedge e_j + a^2 \delta^*_{\cdot} T (1 \leq i \leq d) \sum_{l=1}^{i-1} e_l \wedge e_j + \nonumber \eeq
 \beq + a^2 T (1 \leq j \leq d) \sum_{l=1}^{j-1} e_i \wedge e_l  +    a^2 \sum_{1 \leq l < k \leq d}  e_k \wedge e_i \wedge e_l \wedge e_j + \nonumber \eeq
\beq - e_i \wedge e_j (a^2 d - a^2 \delta^*_{\wedge} T (1 \leq i \leq d) - a^2 T (1 \leq j \leq d)) \eeq 
Thus, we obtain
\beq \int_{\theta_1 \in  \Gamma_{d_1, a_1}, \theta_2 \in \Gamma_{d_2, a_2}} (d \theta_1 * d \theta_2) \cdot (\theta_1 \wedge \theta_2)= \nonumber \eeq
\beq = \frac{a_2^2}{2} \sum_{i=1}^{d_2} \bigg(a_1^2 \delta^*_{\cdot} \sum_{l=1}^{d_1} e_l \wedge e_i - a_1^2 \sum_{k,l \in \{1, \cdots, d_1 \} \setminus \{i \}} e_k \wedge e_l T (l<k)- \frac{a_1^2 d_1 - a_1^2 T(1 \leq i \leq d_1)}{2} \bigg) + \nonumber \eeq
\beq + a_2^2 \sum_{1 \leq j < i \leq d_2} \bigg( a_1^2 T (1 \leq i <j \leq d_1)+ \nonumber\eeq
\beq + a_1^2 T (1 \leq i \leq d_1) \sum_{k=i+1}^{d_1} e_k \wedge e_j + a_1^2 \delta^*_{\cdot} T (1 \leq i \leq d_1) \sum_{l=1}^{i-1} e_l \wedge e_j + \nonumber \eeq
 \beq + a_1^2 T (1 \leq j \leq d_1) \sum_{l=1}^{j-1} e_i \wedge e_l  +    a_1^2 \sum_{1 \leq l < k \leq d_1}  e_k \wedge e_i \wedge e_l \wedge e_j + \nonumber \eeq
\beq - e_i \wedge e_j (a_1^2 d_1 - a_1^2 \delta^*_{\wedge} T (1 \leq i \leq d_1) - a_1^2 T (1 \leq j \leq d_1)) \bigg) \label{Longest} \eeq
Let us now evaluate it term by term. The first term is 
\beq \frac{a_2^2}{2} \sum_{i=1}^{d_2} \bigg(a_1^2 \delta^*_{\cdot} \sum_{l=1}^{d_1} e_l \wedge e_i \bigg) = \frac{a_1^2a_2^2}{2} \delta^*_{\cdot} \sum_{i=1}^{d_2} \sum_{l=1}^{d_1} (e_l \wedge e_i)  \eeq
The second term is 
\beq \frac{a_2^2}{2} \sum_{i=1}^{d_2} \bigg(-a_1^2 \sum_{k,l \in \{1, \cdots, d_1\} \setminus \{i \}} e_k \wedge e_l T(l<k) \bigg) = \nonumber \eeq
\beq = - \frac{a_1^2a_2^2}{2} \sum_{k,l \in \{1, \cdots, d_1 \}} \bigg( e_k \wedge e_l T(l<k) \sum_{i=1}^{d_2} T (i \neq k) T (i \neq l) \bigg) = \nonumber \eeq 
\beq = - \frac{a_1^2 a_2^2}{2} \sum_{k,l \in \{1, \cdots, d_1 \}} \Big(e_k \wedge e_l T (l<k) (d_2-2) \Big)= - \frac{a_1^2a_2^2 (d_2-2)}{2} \sum_{1 \leq l < k \leq d_1} e_k \wedge e_l \eeq
The third term is 
\beq \frac{a_2^2}{2} \sum_{i=1}^{d_2} \bigg(- \frac{a_1^2d_1}{2} \bigg) = - \frac{a_1^2a_2^2d_1}{4} \sum_{i=1}^{d_2} 1 = - \frac{a_1^2a_2^2d_1d_2}{4} \eeq 
The fourth term is 
\beq \frac{a_2^2}{2} \sum_{i=1}^{d_2} \bigg(- \frac{-a_1^2 T(1 \leq i \leq d_1)}{2} \bigg) = \frac{a_1^2a_2^2}{4} \sum_{i=1}^{d_2} T (1 \leq i \leq d_1) = \nonumber \eeq 
\beq = \frac{a_1^2a_2^2}{4} \sum_{i=1}^{\min (d_1,d_2)} 1 = \frac{a_1^2a_2^2}{4} \min (d_1,d_2) \eeq
The fifth term is 
\beq \frac{a_2^2}{2} \sum_{i=1}^{d_2} \bigg(a_2^2 \sum_{1 \leq j <i \leq d_2} (a_1^2 T(1 \leq i <j \leq d_1) ) \bigg) = 0 \eeq
due to the fact that $T (1 \leq i <j \leq d_1)=0$ whenever the condition of the sum, $1 \leq j <i \leq d_2$, is met. The sixth term is
\beq a_2^2 \sum_{1 \leq j < i \leq d_2} \bigg(a_1^2 T (1 \leq i \leq d_1) \sum_{k=i+1}^{d_1} e_k \wedge e_j \bigg)  =a_1^2a_2^2 \sum_{1 \leq j <i \leq \min (d_1,d_2)} \sum_{k=i+1}^{d_1} e_k \wedge e_j = \nonumber \eeq
\beq  =a_1^2a_2^2 \sum_{1 \leq j < \min (d_1,d_2)} \bigg( \sum_{k=i+1}^{d_1} e_k \wedge e_j \sum_{i=j+1}^{\min (d_1,d_2)} 1 \bigg) =  \nonumber \eeq
\beq = a_1^2 a_2^2 \sum_{j=1}^{\min (d_1,d_2)-1} \bigg((\min (d_1,d_2)-j) \sum_{k=j+1}^{d_1} e_k \wedge e_j \bigg) \eeq
The seventh term is 
\beq a_2^2 \sum_{1 \leq j <i \leq d_2} \bigg(a_1^2 \delta^*_{\cdot} T (1 \leq i \leq d_1) \sum_{l=1}^{i-1} e_l \wedge e_j \bigg)= a_1^2 a_2^2 \delta^*_{\cdot} \sum_{1 \leq j <i \leq \min (d_1,d_2)} \sum_{l=1}^{i-1} e_l \wedge e_j = \nonumber \eeq 
\beq = a_1^2 a_2^2 \delta^*_{\cdot} \sum_{j=1}^{\min (d_1,d_2)-1} \; \; \; \sum_{l=1}^{\min (d_1,d_2)-1} \; \; \; \sum_{i= \max (j,l)+1}^{\min (d_1,d_2)} e_l \wedge e_j = \nonumber \eeq
\beq =  a_1^2 a_2^2 \delta^*_{\cdot} \sum_{j=1}^{\min (d_1,d_2)-1} \; \; \; \sum_{l=1}^{\min (d_1,d_2)-1}  (e_l \wedge e_j (\min (d_1,d_2) - \max (j,l)) \eeq
The eighth term is 
\beq a_2^2 \sum_{1 \leq j < i \leq d_2} \bigg(a_1^2 T(1 \leq j \leq d_1) \sum_{l=1}^{j-1} e_i \wedge e_l \bigg) = a_1^2 a_2^2 \sum_{1 \leq l < j < i  \leq d_2} (T (1 \leq j \leq d_1) e_i \wedge e_l) = \nonumber \eeq
\beq = a_1^2 a_2^2 \sum_{1 \leq l <j \leq \min (d_1,i-1) <i \leq d_2} e_i \wedge e_l = a_1^2 a_2^2 \sum_{1 \leq l <\min (d_1,i-1) <i \leq d_2} \sum_{j=l+1}^{\min (d_1, i-1)} e_i \wedge e_l  = \nonumber \eeq 
\beq = a_1^2 a_2^2 \sum_{1 \leq l <\min (d_1,i-1) <i \leq d_2} \Big(\big(\min (d_1, i-1)-l \big) e_i \wedge e_l  \Big) \eeq 
The ninth term is
\beq a_2^2 \sum_{1 \leq j < i \leq d_2} \bigg(a_1^2 \sum_{1 \leq l < k \leq d_1} e_k \wedge e_i \wedge e_l \wedge e_j \bigg) = a_1^2 a_2^2 \sum_{1 \leq j <i \leq d_2} \sum_{1 \leq l < d \leq d_1} e_k \wedge e_i \wedge e_l \wedge e_j \eeq 
The tenth term is 
\beq a_2^2 \sum_{1 \leq j <i \leq d_2} \Big(- e_i \wedge e_j \; a_1^2 d_1 \Big) = - a_1^2 a_2^2 d_1 \sum_{1 \leq j < i \leq d_2} e_i \wedge e_j \eeq
The eleventh term is 
\beq a_2^2 \sum_{1 \leq j < i \leq d_2}  \Big( -e_i \wedge e_j \big(-a_1^2  \delta^*_{\wedge} T (1 \leq i \leq d_1) \big) \Big) = a_1^2 a_2^2 \delta^*_{\wedge} \sum_{1 \leq j < i \leq d_2} \big( e_i \wedge e_j \; T (1 \leq i \leq d_1) \big) = \nonumber \eeq
\beq = a_1^2 a_2^2 \delta^*_{\wedge} \sum_{1 \leq j < i \leq \min (d_1,d_2)} e_i \wedge e_j \eeq
And finally, the twelth term is 
\beq a_2^2 \sum_{1 \leq j < i \leq d_2} \Big( -e_i \wedge e_j \big(-a_1^2 T (1 \leq j \leq d_1) \big) \Big) = a_1^2 a_2^2 \sum_{1 \leq j < i \leq d_2} \big( e_i \wedge e_j T(1 \leq j \leq d_1) \big) = \nonumber \eeq
\beq = a_1^2 a_2^2 \sum_{1 \leq j \leq \min (i-1, d_1) <i \leq d_2} e_i \wedge e_j \eeq
Therefore, after pulling those terms together, we obtain
\beq \int_{\theta_1 \in \Gamma_{d_1,a_1}, \theta_2 \in \Gamma_{d_2,a_2}} (d \theta_1 * d \theta_2) \cdot (\theta_1 \wedge \theta_2) = \nonumber \eeq
\beq = \frac{a_1^2a_2^2}{2} \delta^*_{\cdot} \sum_{i=1}^{d_2} \sum_{l=1}^{d_1} (e_l \wedge e_i) - \frac{a_1^2a_2^2 (d_2-2)}{2} \sum_{1 \leq l < k \leq d_1} e_k \wedge e_l - \frac{a_1^2a_2^2d_1d_2}{4} +  \nonumber \eeq
\beq + \frac{a_1^2a_2^2}{4} \min (d_1,d_2)  +  a_1^2 a_2^2 \sum_{j=1}^{\min (d_1,d_2)-1} \bigg((\min (d_1,d_2)-j) \sum_{k=j+1}^{d_1} e_k \wedge e_j \bigg) + \nonumber \eeq
\beq +  a_1^2 a_2^2 \delta^*_{\cdot} \sum_{j=1}^{\min (d_1,d_2)-1} \; \; \; \sum_{l=1}^{\min (d_1,d_2)-1}  (e_l \wedge e_j (\min (d_1,d_2) - \max (j,l))  \nonumber \eeq
\beq + a_1^2 a_2^2 \sum_{1 \leq l <\min (d_1,i-1) <i \leq d_2} \Big(\big(\min (d_1, i-1)-l \big) e_i \wedge e_l  \Big)  +a_1^2 a_2^2 \sum_{1 \leq j <i \leq d_2} \sum_{1 \leq l < d \leq d_1} e_k \wedge e_i \wedge e_l \wedge e_j  -  \nonumber \eeq
\beq - a_1^2 a_2^2 d_1 \sum_{1 \leq j < i \leq d_2} e_i \wedge e_j + a_1^2 a_2^2 \delta^*_{\wedge} \sum_{1 \leq j < i \leq \min (d_1,d_2)} e_i \wedge e_j +  a_1^2 a_2^2 \sum_{1 \leq j \leq \min (i-1, d_1) <i \leq d_2} e_i \wedge e_j \eeq
Now, if we set 
\beq a_1 = \sqrt{\frac{2}{d_1}} \; , \; a_2 = \sqrt{\frac{2}{d_2}} \eeq
then, by observing that every single term above contains $a_1^2 a_2^2$, we conclude that it contains the factor of 
\beq a_1^2a_2^2 = \frac{4}{d_1d_2} \eeq
Therefore, in order to prevent it from going to zero as $d_1 \rightarrow \infty$ and $d_2 \rightarrow \infty$, we need an extra factor of $d_1d_2$ in the numerator in order to cancel the one in denominator. The only term that contains such a factor is the third term. Therefore, the third term is the only one that survives under the above limit, and we obtain 
\beq \lim^{max}_{d_1 \rightarrow \infty, d_2 \rightarrow \infty} \int_{\theta_1 \in \Gamma_{d_1,a_1}, \theta_2 \in \Gamma_{d_2,a_2}} (d \theta_1 * d \theta_2) \cdot (\theta_1 \wedge \theta_2) = - 1 \label{WrongSingMarch} \eeq
The reason why this is $-1$ rather than $+1$ has been discussed in Section \ref{SignSection}. 

\section{Arbitrary number of iterated integrals}

Let us now discuss a more general integral, of the form 
\beq \int_{\theta_1 \in \Gamma_{d_1,a_1} , \cdots, \theta_N \in \Gamma_{d_N, a_N}} (d \theta_1 * \cdots * d \theta_N) \cdot (\theta_{b_1} \wedge \cdots \wedge \theta_{b_M})  \label{GeneralForm} \eeq 
where we assume that 
\beq \{b_1, \cdots, b_M \} \subset \{1, \cdots, N \} \label{SubsetCondition} \eeq
and, without the loss of the generality (due to the anticommutativity of the wedge product) we further assume that 
\beq b_1 < \cdots < b_M \label{WithoutLossOfGenerality} \eeq
yet it doesn't necessarily match $\{1, \cdots, N \}$ since we skip over some of the variables, as we have done, for example, with $(d \theta_1 * d \theta_2) \cdot \theta_2$. 

As one could see from the case of the single and double integrals, the expressions for finite $a_n$ and $d_n$ were quite complicated (where $n \in \{1,2 \}$ as far as the previous sections are concerned, and $n \in \{1, \cdots, N \}$ in this section); yet the lim-max of $d_n \rightarrow \infty$ with 
\beq a_n = \sqrt{\frac{2}{d_n}} \label{daGeneral} \eeq
returned simple answers of $0$ or $\pm 1$. In other words, the "complications" involve the infinitesimal (in lim-max sense) deviations from $0$ and $\pm 1$, which, at the end of the day, we don't care about. Therefore, in order to spare ourselves from even more complicated work, we will avoid doing the finite calculation for the general case and, instead, simply come up with a hand-waving argument (inspired by the inspection of the previous calculations) that lim-max will return $0$ and $\pm 1$ as desired. 

Suppose we know that $\theta_n \in \Gamma_{a_n, d_n}$ where $a_n$ is given by Eq \ref{daGeneral}. Furthermore, suppose that $d \theta_n$ lies on the edge number $c_n$. Then, we immediately know that 
\beq 1 \leq c_n \leq d_n \eeq
\beq d \theta_n = e_{j_n} dt \label{dGeneral} \eeq
\beq \theta_n = a_n \sum_{i_n=1}^{j_n-1} e_i + t e_{j_n} \label{ThetaGeneral} \eeq 
This means that Eq \ref{GeneralForm} produces superposition of the terms of the form 
\beq K \bigg( \prod_{n=1}^N a_n \bigg) \bigg( \prod_{m=1}^M a_{b_m} \bigg) (e_{j_1} * \cdots * e_{j_N}) \cdot (e_{i_{b_1}} \wedge \cdots \wedge e_{i_{b_M}}) \label{Klara} \eeq 
where the appearance of $a$-s is clear from dimensional analysis combined with the inspection of our earlier calculations, and $K$ is some finite factor obtained from the product of $\pm 1$-s, $\half$-s and other trivial things we dealt with earlier. Now, suppose we are seeking out the term of the form 
\beq e_{k_1} \wedge \cdots \wedge e_{k_L} \eeq 
First of all, there is no way for $L$ to possibly exceed $M+N$. On the other hand, if it happens that $j_n=i_{b_m}$ for some $1 \leq n \leq N$ and $1 \leq m \leq M$ then $e_{j_n}$ and $e_{i_{b_m}}$ will "annihilate" each other when we take a product (with the additional $\pm 1$ coefficient), which would allow $L$ to be less than $M+N$. Now, in order to make $L$ as small as possible, we have to use up every single $e_{i_{b_m}}$ (thus, instead of $M$ of them there will be $0$ of them) in annihilating $e_{j_n}$ (thus, instead of $N$, there would be $N-M$ of them) resulting in a total of $N-M$ remaining $e$-s. Thus, we conclude
\beq N-M \leq L \leq N+M \label{ConditionOnL} \eeq 
Since $L$ can only decrease in pairs, we also know that 
\beq N+M-L = {\rm Even} \eeq
and, equivalently, 
\beq L- (N-M) = {\rm Even} \eeq
 The total number of contracted pairs is given by 
\beq \sharp \{ {\rm contracted \; pairs} \} = \frac{N+M-L}{2} \eeq
The Eq \ref{ConditionOnL} implies
\beq 0 \leq \frac{N+M-L}{2} \leq M \leq N \label{ConditionOnContractionNumber} \eeq
Now, if we know that $n$ and $b_m$, then we know that $j_n$ is bounded by $d_n$ and $i_{b_m}$ is bounded by $d_{b_m}$. But, in order for $e_{j_n}$ and $e_{i_{b_m}}$ to contract, we need to have $j_n=i_{b_m}$. Thus, they are bounded by the common upper bound $\min (d_n, d_{b_m})$. We then have to take the product of $(N+M-L)/2$ different pairs $(m_l,n_l)$ and also sum over all possible choices of $m_l \in \{1, \cdots, M \}$ and $n_l \in \{1, \cdots, N \}$. Thus, the combinatoric factor takes the form 
\beq {\rm Combinatoric \; factor} = C \sum_{m_1=1}^M \cdots \sum_{m_L=1}^M \sum_{n_1=1}^N \cdots \sum_{n_L=1}^N \prod_{l=1}^{\frac{N+M-L}{2}} \min (d_{n_l}, d_{b_{m_l}}) \label{Clara} \eeq 
where, due to the fact that we get some $\pm$ signs that we have not taken into account, we would expect
\beq -1 \leq C \leq 1 \eeq 
By combining Eq \ref{daGeneral}, \ref{Klara} and \ref{Clara}, we obtain 
\beq  CK \sum_{m_1=1}^M \cdots \sum_{m_L=1}^M \sum_{n_1=1}^N \cdots \sum_{n_L=1}^N \bigg[ \bigg( \prod_{n=1}^N \sqrt{\frac{2}{d_n}} \bigg) \bigg( \prod_{m=1}^M \sqrt{\frac{2}{d_{b_m}}} \bigg) \bigg( \prod_{l=1}^{\frac{N+M-L}{2}}  \min (d_{n_l}, d_{b_{m_l}}) \bigg)  e_{k_1} \wedge \cdots \wedge e_{k_L} \bigg] \label{CK} \eeq
where $e_{k_1} \wedge \cdots \wedge e_{k_L}$ is some afore-given product we have in mind, and we are counting all the possible ways of obtaining it. Now, from Eq \ref{ConditionOnContractionNumber} we know that 
\beq \prod_{n=1}^N d_n = \bigg( \prod_{l=1}^{\frac{N+M-L}{2}} \min (d_1, d_2) \bigg) \prod \{ {\rm Other \; d_n's} \} \label{Other1} \eeq
\beq \prod_{n=1}^M d_{b_m} = \bigg( \prod_{l=1}^{\frac{N+M-L}{2}} \min (d_1, d_2) \bigg) \prod \{ {\rm Other \; d_{b_m}'s} \} \label{Other2} \eeq
where we have assumed the "most likely" situation that 
\beq {\rm Most \; Likely} \Longrightarrow \min (d_{n_l}, d_{b_{m_l}}) \neq \min (d_{n_{l'}}, d_{b_{m_{l'}}}) \eeq
after we will convince the reader that some other things go to zero as $d_n \rightarrow \infty$, the reader will hopefully be able to also convince himself that the contributions of the "less likely" situations we are neglecting go to zero as well. Anyway, from Eq \ref{Other1} and \ref{Other2}, the Eq \ref{CK} becomes 
\beq   CK \sum_{m_1=1}^M \cdots \sum_{m_L=1}^M \sum_{n_1=1}^N \cdots \sum_{n_L=1}^N  \sqrt{\frac{2^{N+M}}{\prod \{ {\rm Other \; d's} \}}} \label{NeedToGetRidOfOther0} \eeq 
where
\beq {\rm Most \; Likely} \Longrightarrow \{ {\rm Other \; d's} \} = \{ {\rm Other \; d_n's} \} \cup \{ {\rm Other \; d_{b_m}'s} \} \eeq
Now, it is easy to see that 
\beq {\rm Most \; Likely} \Longrightarrow \sharp \{ {\rm Other \; d_n's} \} = N - \frac{N+M-L}{2} \eeq 
\beq {\rm Most \; Likely} \Longrightarrow \sharp \{ {\rm Other \; d_{b_m}'s} \} = M - \frac{N+M-L}{2} \eeq 
and, therefore, 
\beq {\rm Most \; Likely} \Longrightarrow \sharp \{ {\rm Other \; d's} \} = \bigg( N - \frac{N+M-L}{2} \bigg) + \bigg( M - \frac{N+M-L}{2} \bigg)=L \eeq 
But the right hand side of Eq \ref{NeedToGetRidOfOther0} tells us that, as long as the number of "other $d$-s" is non-zero, the answer will go to zero as $d \rightarrow \infty$. Therefore, 
\beq {\rm LimMax \; Doesn't \; Approach \; 0} \Longrightarrow \sharp \{ { \rm Other \; d's} \} = 0 \Longrightarrow L=0 \eeq
In other words, the only part with a non-zero coefficient in the limit of $d_k \rightarrow \infty$ is the scalar. Indeed, in the previous sections we have seen that, after taking the lim-max, we were left with the scalar term. It is important to note that this only applies to the lim-max and not to the regular limit. After all, the above argument shows that the coefficient next to the \emph{particular} $e_{k_1} \wedge \cdots \wedge e_{k_L}$. Now, if we were to have the Pythagorean metric, it would take the form 
\beq {\rm Pythagorean} = \sqrt{\sum_{k=1}^{P} \epsilon^2} = \epsilon \sqrt{P} \eeq
where $P$ is the total number of selections of $e_{k_1} \wedge \cdots \wedge e_{k_L}$, given by 
\beq L \ll d_k \Longrightarrow P \approx \prod_{l=1}^L d_{k_l} \eeq 
which means that the increase in magnitude due to the multiplication by $\sqrt{P}$ can, at least in principle, decrease in magnitude due to the division by the square roots of "other $d$-s". On the other hand, if we are dealing with the max-norm instead of the Pythagorean-norm, then the value of $P$ becomes irrelevant, and the max-norm remains $\epsilon$ rather than $\epsilon \sqrt{P}$. So, in this case, all that matters is that $\epsilon$ is small, and the latter is the case due to the presence of the "other $d$-s" which is linked to $L \neq 0$. Thus, it is strictly the max-norm that tells us that $L \neq 0$ cases approach zero. 

Anyway, now that we have established that $L=0$, it is easy to see that this can be accomplished only with $M=N$. To check if that's the case, one could use Eq \ref{ConditionOnContractionNumber}: 
\beq L=0 \Longrightarrow^{Eq \; \ref{ConditionOnContractionNumber}} \frac{N+M}{2} \leq M \Longrightarrow \frac{N+M}{2} \leq \frac{M+M}{2} \Longrightarrow N \leq M \eeq
yet, at the same time
\beq {\rm Eq \; \ref{ConditionOnContractionNumber}} \Longrightarrow M \leq  N \eeq
and, therefore
\beq L=0 \Longrightarrow M=N \eeq 
But Eq \ref{SubsetCondition} implies that 
\beq M=N \Longrightarrow \{b_1, \cdots, b_M \} = \{1, \cdots, N \} \eeq
Thus, the only non-zero integrals are permutations of 
\beq \int (d \theta_1 * \cdots *d \theta_N) \cdot (\theta_1 \wedge \cdots \wedge \theta_N) \label{DidntPermute} \eeq
Since "in most cases" $d \theta_i$ and $d \theta_j$ occupy a different edge, "in most cases" the star-product coincides with the wedge-product and anticommutes. Since in the lim-max only "most cases" survive, the permutation of differentials simply changes the sign as far as the lim-max is concerned. Thus, we can make things simpler and just look at Eq \ref{DidntPermute} without worrying about its permutations. Now, the above expression will produce products of the form 
\beq (e_{j_1} * \cdots * e_{j_N}) \cdot (e_{i_1} \wedge \cdots \wedge e_{i_N}) \eeq 
Now, if we plug in Eq \ref{dGeneral} and \ref{ThetaGeneral} into Eq \ref{DidntPermute}, it is clear that we would have 
\beq \forall n \in \{1, \cdots, N \} ( i_n \leq j_n) \label{GeneraljLessThani} \eeq
Now suppose $(k_1, \cdots, k_N )$ is a re-ordering of $(1, \cdots, N)$ such that 
\beq i_{k_1} < \cdots < i_{k_N} \label{GeneralOrderk} \eeq
where we know that $i_k \neq i_l$ because $\theta_{i_k} \wedge \theta_{i_l} \neq 0$. Now, as we established earlier, only scalar survives the limit. But, in order to have a scalar, each $e_{j_n}$ has to be contracted with some $e_{i_m}$. Now, the combination of Eq \ref{GeneraljLessThani} and \ref{GeneralOrderk} tells us that 
\beq n \geq 2 \Longrightarrow i_{k_1} < i_{k_n} \leq j_{k_n} \eeq 
and, therefore, $i_{k_1}$ cannot be contracted with $j_{k_n}$. But $i_{k_1}$ has to be contracted with \emph{something}. So the fact that it can't be contracted with $j_{k_n}$ for $n \geq 2$ implies that it should be contracted with $j_{k_1}$. But the latter contraction requires 
\beq i_{k_1} = j_{k_1} \eeq
We would now like to find what to couple $i_{k_2}$ to. Again, the combination of Eq \ref{GeneraljLessThani} and \ref{GeneralOrderk} tells us 
\beq n \geq 3 \Longrightarrow i_{k_2} < i_{k_n} \leq j_{k_n} \eeq
so $i_{k_2}$ can't be contracted with $j_{k_n}$ for $n \geq 3$. Furthermore, $i_{k_2}$ can't be contracted with $j_{k_1}$ since the latter has already been contracted with $i_{k_1}$. Therefore, the only thing $i_{k_2}$ can be contracted with is $j_{k_2}$. So, in order to get a scalar, we have no choice but to contract them, which means
\beq i_{k_2}= j_{k_2} \eeq
As we keep going in the same fashion, we can show by induction that 
\beq {\rm Scalar} \Longrightarrow \forall n \in \{1, \cdots, N \} (i_{k_n}=j_{k_n}) \eeq 
But we know that $(k_1, \cdots, k_N)$ is merely a re-ordering of $(1, \cdots, N)$. Thus, we conclude that 
\beq {\rm Scalar} \Longrightarrow \forall n \in \{1, \cdots, N \} (i_n = j_n) \eeq 
and, since only the scalar survives the lim-max, we have 
\beq \lim^{max} \neq 0 \Longrightarrow {\rm Scalar} \Longrightarrow \forall n \in \{1, \cdots, N \} (i_n=j_n) \eeq 
Now, as one can readily see by inspecting some of our derivations of single and double integrals, same-index contraction produces the coefficient of 
\beq \int_0^a tdt = \frac{a^2}{2} \eeq
while different-index contraction produces 
\beq a \int_0^a dt = a^2 \eeq
 Since now we have $N$ same-index contractions, we have 
\beq \prod_{n=1}^N \frac{a_n^2}{2} \label{SingeEdgeGeneral} \eeq
However, as the above integrals indicate, they are only taken over a single edge. So now we have to multiply by all possible choices of edges. In other words, we have to multiply by the number of choices of $(i_1, \cdots, i_N)$ (and we don't have to count the number of $j$-s since we have already established that $j_n=i_n$) Now the condition that $i_n \neq i_{n'}$ implies that, once we fill some of the slots, we have fewer and fewer options. However, this won't have a significant effect \emph{if} we assume $N \ll \min (d_1, \cdots, d_N)$; in other words, the dimensionalities of the contours are much greater than the number of integral signs (which is self evident since the former is sent to infinity while the latter stays fixed). Thus, 
\beq N \ll \min (d_1, \cdots, d_N) \Longrightarrow \sharp \{ {\rm Edge \; Combinations} \} \approx \prod_{k=1}^N d_n \label{NumberEdgeCombinationsGeneral}\eeq
The combination of Eq \ref{SingeEdgeGeneral} and \ref{NumberEdgeCombinationsGeneral} implies that
\beq N \ll \min (d_1, \cdots, d_N) \Longrightarrow \nonumber \eeq
\beq \Longrightarrow  \lim_{d_1 \rightarrow \infty \cdots d_N \rightarrow \infty}^{max} \int_{\theta_1 \in \Gamma_{d_1,a_1}, \cdots, \theta_N \in \Gamma_{d_N,a_N}} (d \theta_1 * \cdots * d \theta_N) \cdot (\theta_1 \wedge \cdots \wedge \theta_N) = \pm \prod_{n=1}^N \frac{da_n^2}{2} = \pm 1 \eeq
where, in the last step, we assumed that 
\beq a_n = \sqrt{\frac{2}{d_n}} \eeq
and we put an exact sign rather than approximation because we were taking the LimMax; the proof that the LimMax is indeed exact can be understood intuitively upon close inspection of the combinatorial aspects of the various calculations that were presented; the rigourous proof is beyond the scope of this paper. The sign of $\pm 1$ comes from the need of rearranging $e$-s in order to obtain contractions. For example, 
\beq (e_1 \wedge e_2) \cdot (e_1 \wedge e_2) = - (e_1 \wedge e_2) \cdot (e_2 \wedge e_1) = \nonumber \eeq
\beq =  - (e_1 \cdot e_2) \cdot (e_2 \cdot e_1) = - e_1 \cdot (e_2 \cdot e_2) \cdot e_1 = - e_1 \cdot 1 \cdot e_1 = - e_1 \cdot e_1 = -1 \eeq 
The author is well aware that, conventionally, this integral is taken to be $+1$; the difference between our conventions and the more standard conventions is discussed in Section \ref{SignSection}. In any case, we first move $\theta_N$ to the left, which requires $(-1)^{N-1}$, then we move $\theta_{N-1}$ to the left which requires $(-1)^{N-2}$, and so forth. Thus, the total factor is 
\beq \prod_{n=1}^N (-1)^{n-1} = (-1)^{\sum_{n=1}^N (n-1)} = (-1)^{N(N-1)/2} \eeq 
and, therefore, 
\beq N \ll \min (d_1, \cdots, d_N) \Longrightarrow \nonumber \eeq
\beq \Longrightarrow  \lim_{d_1 \rightarrow \infty \cdots d_N \rightarrow \infty}^{max} \int_{\theta_1 \in \Gamma_{d_1,a_1}, \cdots, \theta_N \in \Gamma_{d_N,a_N}} (d \theta_1 * \cdots * d \theta_N) \cdot (\theta_1 \wedge \cdots \wedge \theta_N) =  (-1)^{N(N-1)/2}\eeq
Now we would like to know what happens if we multiply the integrand by  some anticommutting constants. If we multiply it by $e_{p_1} \wedge \cdots \wedge e_{p_q}$, then, \emph{provided that} $q \ll \min (d_1, \cdots, d_N)$, in the majority of cases, the edges that $\theta$-s and $d \theta$-s select do \emph{not} coincide with $e_{p_r}$. As a result, $e_{p_1} \wedge \cdots \wedge e_{p_q}$ simply comes along for the ride. Roughly speaking, it works via the following scheme:
\beq 1 \ll \min (d_1, \cdots, d_N) \Longrightarrow {\rm C \; Doesnt \: Overlap} \Longrightarrow  \nonumber \eeq
\beq \Longrightarrow X \cdot C = X \wedge C \Longrightarrow A \cdot (B \wedge C) = A \cdot (B \cdot C) = (A \cdot B) \cdot C = (A \cdot B) \wedge C \eeq 
where 
\beq A = d \theta_1 * \cdots * d \theta_N \eeq
\beq B = \theta_1 \wedge \cdots \wedge \theta_M \eeq
\beq C = e_{p_1} \wedge \cdots \wedge e_{p_q} \eeq 
Thus, by using the $0$-s and $\pm 1$-s we just discussed, we conclude that 
 \beq M=N \Longrightarrow \nonumber \eeq
\beq \Longrightarrow \lim_{d_1 \rightarrow \infty \cdots d_N \rightarrow \infty}^{max} \int_{\theta_1 \in \Gamma_{d_1,a_1}, \cdots, \theta_N \in \Gamma_{d_N,a_N}} (d \theta_1 * \cdots * d \theta_N) \cdot (\theta_1 \wedge \cdots \wedge \theta_M \wedge e_{p_1} \wedge \cdots \wedge e_{p_q}) = \nonumber \eeq
\beq =   (-1)^{N(N-1)/2} e_{p_1} \wedge \cdots \wedge e_{p_q}\eeq

 \beq M<N \Longrightarrow \eeq
\beq \Longrightarrow \lim_{d_1 \rightarrow \infty \cdots d_N \rightarrow \infty}^{max} \int_{\theta_1 \in \Gamma_{d_1,a_1}, \cdots, \theta_N \in \Gamma_{d_N,a_N}}  (d \theta_1 * \cdots * d \theta_M) \cdot (\theta_1 \wedge \cdots \wedge \theta_M \wedge e_{p_1} \wedge \cdots \wedge e_{p_q}) = 0  \nonumber\eeq

Now, any constant $\eta$ is a superposition of $e$-s. Thus, by linearity, we read off 

 \beq M=N \Longrightarrow \nonumber \eeq
\beq \Longrightarrow \lim_{d_1 \rightarrow \infty \cdots d_N \rightarrow \infty}^{max} \int_{\theta_1 \in \Gamma_{d_1,a_1}, \cdots, \theta_N \in \Gamma_{d_N,a_N}} (d \theta_1 * \cdots * d \theta_N) \cdot (\theta_1 \wedge \cdots \wedge \theta_M \wedge \eta_1 \wedge \cdots \wedge \eta_q) = \nonumber \eeq
\beq =   (-1)^{N(N-1)/2} \eta_1 \wedge \cdots \wedge \eta_q \eeq

 \beq M<N \Longrightarrow \eeq
\beq \Longrightarrow \lim_{d_1 \rightarrow \infty \cdots d_N \rightarrow \infty}^{max} \int_{\theta_1 \in \Gamma_{d_1,a_1}, \cdots, \theta_N \in \Gamma_{d_N,a_N}}  (d \theta_1 * \cdots * d \theta_M) \cdot (\theta_1 \wedge \cdots \wedge \theta_M \wedge \eta_1 \wedge \cdots \wedge \eta_q) = 0  \nonumber\eeq
In all of the above statements, the assumption that $q \ll \min (d_1, \cdots, d_N)$ was crucial in order for us to be able to assume that $e$-s coming from constants don't overlap with $e$-s coming from variables. At the same time, such an assumption becomes self-evident when we take a limit of $d_n \rightarrow \infty$, which we did. 

\subsection*{Analytic and non-analytic exponentials} \label{NonAnalyticExponential}

As we have said earlier, in the typical situation the dot-product is reserved strictly for the differential parts, while in finite part of the integral the wedge product is exclusively used. However, the integral will still remain well defined if we replace wedge product with a dot product in a finite part. The only "problem" is that its well defined value would no longer match the conventional one. This, however, is perfectly fine: we simply remember that we have to use wedge product in the finite part to get conventional result, while the dot product is something that we can use if our intention is to go outside of conventional realm. As an example, since conventional QFT doesn't deal with the problem of measurement, the latter can be a good excuse to be unconventional. In particular, according to GRW model, measurement is due to the multiplication of wave function by Gaussians. \emph{But} it is easy to see that wedge-based Gaussian of anticommutting number is a constant and, therefore, accomplishes nothing. On the other hand, if we use dot-based Gaussian we would, in fact, get non-trivial result. Be it as it may, the integral of dot-based Gaussian would still give zero (see Eq {RealFunctionZero}). At the same time, the integral of some \emph{other} function \emph{would}, in fact, change after its multiplication by Gaussian (see {UsualIntegralsFinally}) and, therefore, GRW model would be of some consequence. In any case, as far as this paper is concerned, it is not our intention to talk about quantum measurement models. Therefore, we will limit ourselves to the integration of non-analytic exponential just to get the concept of non-analytic integrals across, which would then be used in future papers. 

 The star-product between two Grassmann numbers computes as 
\beq \bigg( \sum_k a_k e_k \bigg) * \bigg( \sum_l b_l e_l \bigg) = \sum_{kl} a_k b_l e_k * e_l  = \sum_{k<l} a_k b_l e_k * e_l + \sum_{l<k} a_k b_l e_k * e_l + \sum_k a_k b_l e_k * e_l = \nonumber \eeq
\beq = \sum_{k<l} a_k b_l e_k \wedge e_l + \sum_{l<k} a_k b_l e_k \wedge e_l  + \delta^*_{\cdot} \sum_k a_kb_l =  \sum_{k<l} a_k b_l e_k \wedge e_l - \sum_{l<k} a_k b_l e_l \wedge e_k + \delta^*_{\cdot} \sum_k a_k b_l = \nonumber \eeq 
\beq =  \sum_{k<l} a_k b_l e_k \wedge e_l - \sum_{k<l} a_l b_k e_k \wedge e_l + \delta^*_{\cdot} \sum_k a_kb_l = \sum_{k<l} (a_kb_l-a_lb_k) e_k \wedge e_l + \delta^*_{\cdot} \sum_k a_kb_l \eeq 
Now, we will define (non-analytic) power as 
\beq \theta^{*0} =1 \; , \; \theta^{* (n+1)} = \theta* \theta^{*n} \eeq
and we will also distinguish between two different norms: max-norm and Eucledian norm. If we set
\beq \theta= \sum_k a_k e_k \eeq
then 
\beq \vert \theta \vert_{max} = \bigg\vert \sum_k a_k e_k \bigg\vert_{max} = \max \{a_k \vert k \in \mathbb{N} \} \eeq
\beq \vert \theta \vert_{Euc} = \bigg\vert \sum_k a_k e_k \bigg\vert_{Euc} = \bigg( \sum_k a_k^2 \bigg)^{1/2} \eeq
We then obtain 
\beq \theta * \theta = \delta^*_{\cdot} \sum_k a_k^2 = \delta^*_{\cdot} \vert \theta \vert^2_{Euc} \eeq
which implies that
\beq \theta^{*(2n)} = (\theta* \theta)^n = (\delta^*_{\cdot} \vert \theta \vert^2_{Euc})^n = \delta^n_0 \delta^*_{\cdot} \vert \theta \vert^{2n}_{Euc} \eeq
\beq \theta^{*(2n+1)} = \theta* \theta^{*(2n)} = \delta^n_0 \delta^*_{\cdot} \theta \vert \theta \vert_{Euc}^{2n} \eeq
We then define (non-analytic) exponential as 
\beq \exp_* \theta = \sum_{n=0}^{\infty}  \frac{\theta^{*n}}{n!}  = \sum_{n=0}^{\infty}  \frac{\theta^{*(2n)}}{(2n)!} +  \sum_{n=0}^{\infty}  \frac{\theta^{*(2n+1)}}{(2n+1)!} = \sum_{n=0}^{\infty} \frac{\delta^*_{\cdot} \vert \theta \vert_{Euc}^{2n}}{(2n)!} +  \sum_{n=0}^{\infty} \frac{\delta^*_{\cdot} \theta \vert \theta \vert_{Euc}^{2n}}{(2n+1)!} = \nonumber \eeq
\beq =  \delta^*_{\wedge} (1 + \theta) + \delta^*_{\cdot} \bigg( \cosh \vert \theta \vert_{Euc} + \frac{ \theta}{\vert \theta \vert_{Euc}} \sinh \vert \theta \vert_{Euc} \bigg) \label{ExpStar} \eeq
Now, we already know how to integrate the first two terms on right hand side. Let us evaluate the integrals of last two terms. Suppose $f$ is a real valued function. In other words, even though $\theta$ is Grassmannian, $f (\theta) \in \mathbb{R}$ is real. Then the third term can be evaluated via 
\beq f(\theta) \in \mathbb{R} \cap [-A,A] \Longrightarrow \bigg\vert \int_{\Gamma_{d,a}}  d \theta f (\theta) \bigg\vert \leq A \bigg\vert \int_{\Gamma_{d,a}} d \theta \bigg\vert \Longrightarrow \lim_{d \rightarrow \infty}^{\max} \int_{\Gamma_{d,\sqrt{2/d}}} d \theta f (\theta) = 0 \label{RealFunctionZero} \eeq
Now, in order to evaluate the fourth term, let us assume that 
\beq \epsilon < \sqrt{\frac{2}{d}} \Longrightarrow f (\theta + \epsilon e_k) \approx f (\theta) \eeq
then, if we parametrize $\Gamma_{d, \sqrt{2/d}}$ with a parameter $t$ satisfying 
\beq 0 \leq t \leq 2 \eeq
through 
\beq 0 \leq k \leq d \Longrightarrow \theta \bigg(\frac{2(k-1)}{d}+ t \bigg) = \sqrt{\frac{2}{d}} \sum_{l=1}^{k-1} e_l + te_k \eeq 
 we have 
\beq \lim^{max}_{d \rightarrow \infty} \int_{\Gamma_{d,\sqrt{2/d}}} d \theta \cdot \theta f (\theta) = \lim_{d \rightarrow \infty} \sum_{k=1}^d \bigg( f \bigg(\frac{2k}{d}\bigg) \int_0^{\sqrt{2/d}}  \bigg( (e_k dt) \cdot \bigg ( \sqrt{\frac{2}{d}} \sum_{l=1}^{k-1} e_l + te_k \bigg)\bigg) \bigg) = \nonumber \eeq
\beq = \lim^{max}_{d \rightarrow \infty} \sum_{1 \leq l <k \leq d} \bigg( \sqrt{\frac{2}{d}}  f \bigg(\frac{2k}{d}\bigg)  e_k \cdot e_l \int_0^{\sqrt{2/d}} dt \bigg) + \sum_{k=1}^d \bigg(  f \bigg(\frac{2k}{d}\bigg)   e_k \cdot e_k \int_0^{\sqrt{2/d}} t dt \bigg) = \nonumber \eeq
\beq = \lim^{max}_{d \rightarrow \infty} \sum_{1 \leq l < k \leq d} \bigg( \sqrt{\frac{2}{d}} \sqrt{\frac{2}{d}}  f \bigg(\frac{2k}{d}\bigg)   e_k \wedge e_l) + \sum_{k=1}^d \bigg( \frac{1}{2} \bigg(\sqrt{\frac{2}{d}} \bigg)^2  f \bigg(\frac{2k}{d}\bigg)  \bigg) = \nonumber \eeq
\beq = \lim^{max}_{d \rightarrow \infty} \sum_{1 \leq l < k \leq d} \bigg(\frac{2}{d}  f \bigg(\frac{2k}{d}\bigg)   e_k \wedge e_l) + \sum_{k=1}^d \bigg( \frac{1}{d} f \bigg(\frac{2k}{d}\bigg)  \bigg) = \frac{1}{2} \int_0^2 f(t) dt  \label{UsualIntegralsFinally} \eeq 
where we have sent the first term to zero via LimMax while the second term was sent to \emph{half} the integral, given that 
\beq \frac{1}{t} = \frac{1}{2} \delta t \eeq 
and the upper limit of integration is $2$ due to 
\beq t_{max}= \sum_{k=1}^d (\delta t)_k = d \frac{2}{d} = 2 \eeq 
Now, it is easy to see that
\beq \vert \theta (t) \vert^2_{Euc} = t \eeq
we conclude that 
\beq \lim^{max}_{d \rightarrow \infty} \int_{\Gamma_{d, \sqrt{2/d}}} d \theta \cdot \frac{\theta}{\vert \theta \vert_{Euc}} \sinh \vert \theta \vert_{Euc} = \frac{1}{2} \int_0^2 \frac{\sinh \sqrt{t}}{\sqrt{t}} dt = \nonumber \eeq
\beq = \int_0^2 \sinh \sqrt{t} d \sqrt{t} = \cosh \sqrt{t} \Big\vert_0^2 = \cosh \sqrt{2} - 1 \label{HyperbolicSurprise} \eeq 
If we now plug in Eq \ref{OneVariableConstant}, \ref{RightContourDemo}, \ref{RealFunctionZero} and Eq \ref{HyperbolicSurprise} into Eq \ref{ExpStar} we obtain 
\beq \lim_{d \rightarrow \infty} \int_{\Gamma_{d, \sqrt{2/d}}} d \theta \cdot \exp_* \theta = \delta^*_{\wedge} + \delta^*_{\cdot} \Big(\cosh \sqrt{2}-1 \Big) \eeq
Therefore, in "analytic" case of $* = \wedge$ we get a conventional answer of $1$ and, in "non-analytic" case of $*= \cdot$ we get a "new prediction" involving hyperbolic cosine, which otherwise isn't defined. 

Now, going back to what we talked about earlier, we would like to see how our result changes if we multiply it by (non-analytic) Gaussian. In light of the fact that $\theta* \theta \in \mathbb{R}$, in particular, 
\beq \theta * \theta = \delta^*_{\cdot} \vert \theta \vert_{Euc} \eeq
the equation for non-analytic Gaussian is 
\beq G_* (\theta) = e^{- (\delta^*_{\cdot} \vert \theta \vert_{Euc})^2/2} = \delta^*_{\wedge} + \delta^*_{\cdot}  e^{- \vert \theta \vert_{Euc}^2/2 \vert} \eeq 
as a result, Eq \ref{HyperbolicSurprise} gets modified as 
\beq \lim_{d \rightarrow \infty} \int_{\Gamma_{d, \sqrt{2/d}}} d \theta \cdot  G_* (\alpha \theta) \exp_{**} \theta  = \int_0^2 (\delta^*_{\wedge} + \delta^*_{\cdot} e^{-\alpha^2 t/2} ) \sinh \sqrt{t} d \sqrt{t} \eeq
where we had $e^{- \alpha^2 t/2}$ instead of $e^{- \alpha^2t^2/2}$ due to the fact that $t = \vert \theta_{Euc} \vert^2$. 

\section{Other non-analytic functions} \label{NonAnalytic}

Let us now go a step further and point out that the non-analytic function doesn't have to be expressible in terms of dot-product either. Just about \emph{any} function can be integrated. To get this concept across, let us consider a couple of examples, that don't have any physical motivation as far as the author is aware. In both cases, we will use the same contour we were usually using, and take the same limit, except that we will plug in unusual functions. Consider
\beq f \bigg( \sum x_k e_k \bigg) = \sum x_{k+1} e_k \eeq 
then the integral evaluates to 
\beq \int_{\Gamma_{d,a}} d \theta \cdot f (\theta) = \sum_{k=1}^d \int_0^a (e_k dt) \cdot \bigg(a \sum_{l=1}^{k-2} e_l + te_{k-1} T (k \geq 2) \bigg) = \nonumber \eeq
\beq = \sum_{k=1}^d e_k  \cdot \bigg(a \sum_{l=1}^{k-2} e_l  \int_0^a dt + e_{k-1} T (k \geq 2) \int_0^a tdt \bigg) = \nonumber \eeq 
\beq = \sum_{k=1}^d e_k \cdot \bigg(a e_l a + e_{k-1} \frac{a^2}{2} \bigg) =  a^2  \bigg( \sum_{1 \leq l \leq k-2 <k \leq d} e_k \cdot e_l + \sum_{k=2}^d e_{k-1} \bigg) \nonumber \eeq 
\beq =  a^2  \bigg( \sum_{1 \leq l \leq k-2 <k \leq d} e_k \wedge e_l + \sum_{k=2}^d e_{k-1} \bigg) \eeq
which, in the limit becomes
\beq \lim^{max}_{d \rightarrow \infty} \int_{\Gamma_{d, \sqrt{2/d}}} d \theta \cdot f(\theta) = 0 \eeq
Now lets "shift" the coordinates in the opposite direction:
\beq g  \bigg( \sum x_k e_k \bigg) = \sum x_k e_{k+1} \eeq 
in this case the integral evaluates to 
\beq \int_{\Gamma_{d,a}} d \theta \cdot g (\theta) = \sum_{k=1}^d \int_0^a (e_k dt) \cdot \bigg(a \sum_{l=2}^k e_l + t e_{k+1} \bigg) = \nonumber \eeq
\beq = \sum_{k=1}^d e_k \cdot \bigg(a \sum_{l=1}^k e_l \int_0^a dt + e_{k+1} \int_0^a t dt \bigg) = \nonumber \eeq 
\beq = \sum_{k=1}^d e_k \cdot \bigg(a \sum_{l=1}^k e_l a + e_{k+1} \frac{a^2}{2} \bigg) = a^2 \bigg( \sum_{1 \leq l \leq k \leq d} e_k \cdot e_l + \frac{a^2}{2} e_k \cdot e_{k+1} \bigg) =  \nonumber \eeq 
\beq = a^2 \bigg( \sum_{1 \leq l <k \leq d} e_k \cdot e_l + \sum_{1 \leq k \leq d} e_k \cdot e_k + \frac{a^2}{2} e_k \cdot e_{k+1} \bigg) = \nonumber \eeq 
\beq = a^2 \bigg( \sum_{1 \leq l <k \leq d} e_k \wedge e_l + \sum_{1 \leq k \leq d} 1 + \frac{a^2}{2} e_k \cdot e_{k+1} \bigg) = \nonumber \eeq 
\beq = a^2 \bigg( \sum_{1 \leq l <k \leq d} e_k \wedge e_l + d + \frac{a^2}{2} e_k \cdot e_{k+1} \bigg)  \eeq 
and, therefore, 
\beq \lim^{max}_{d \rightarrow \infty} \int_{\Gamma_{d, \sqrt{2/d}}} d \theta \cdot g (\theta) = 2 \eeq 
Notably, we just obtained $2$, which we never obtained from the analytic integrals (unless, of course, there was an outside coefficient that happened to  be equal to $2$ or an unusual contour was selected, neither of which is the case right now). 

\section{Derivatives} \label{Derivatives}

Let us now turn to a much simpler issue and attempt to define the derivatives with respect to the Grassmann coordinates. The only obstacle to overcome is the fact that we have to ''divide'' by ''vectors''. We propose to define the division as
\beq \frac{A}{\theta} = \frac{\theta \cdot A }{\theta \cdot \theta} \eeq
It then can be easily shown that 
\beq \frac{A}{\theta} = B \Longleftrightarrow \theta \cdot B = A \eeq
via the following calculation: 
\beq \theta \cdot \frac{\theta \cdot A }{\theta \cdot \theta} = \frac{1}{\theta \cdot \theta} \theta \cdot (\theta \cdot A) = \frac{1}{\theta \cdot \theta} (\theta \cdot \theta) \cdot A = A \eeq
where on the last step we were using the assumption that 
\beq k \in \mathbb{C} \Rightarrow \forall A (k \wedge A = A \wedge k = k \cdot A = A \cdot k = kA) \eeq
To write it more explicitly, 
\beq \theta = \sum_k x_k e_k \Longrightarrow \theta \cdot \theta = \bigg(\sum_k x_k e_k \bigg) \cdot \bigg( \sum_l x_l e_l \bigg) = \sum_{kl} x_k x_l e_k \cdot e_l = \nonumber \eeq
\beq= \sum_{kl} x_k x_l (e_k \wedge e_l + \delta^k_l) = \sum_{kl} x_k x_l \delta^k_l = \sum_k x_k^2 \eeq 
and, therefore
\beq \frac{A}{x_1e_1 +x_2 e_2 + \cdots} = \frac{x_1 e_1 \cdot A + x_2 e_2 \cdot A + \cdots}{x_1^2 + x_2^2 + \cdots} \eeq 
It should be noted that if $G \cdot G$ is not real, then division by $G$ is not well defined:  for example, 
\beq (1 + e_1) \cdot (1+e_1) = 1 + 2 e_1 \Longrightarrow \frac{1}{1+e_1} {\rm \; Not \; Defined} \eeq
which is fine with us since the only reason we need the ratios to begin with is to define the derivative, and all of the ratios that occur in the derivative are well-defined based on our definition. In light of the fact that $\theta$ lives in a multidimensional space, we have to define \emph{partial} derivatives as 
\beq \frac{\partial f(\theta)}{\partial \theta_k} = \lim_{\epsilon \rightarrow 0} \frac{f (\theta + \epsilon e_k) - f (\theta)}{\epsilon e_k} = \lim_{\epsilon \rightarrow 0} \frac{\epsilon e_k \cdot (f (\theta + \epsilon e_k) - f (\theta))}{(\epsilon e_k) \cdot (\epsilon e_k)} = \nonumber \eeq
\beq = \lim_{\epsilon \rightarrow 0} \frac{\epsilon e_k \cdot (f (\theta+ \epsilon e_k) - f(\theta))}{\epsilon^2} =  \lim_{\epsilon \rightarrow 0} \frac{e_k \cdot (f (\theta+ \epsilon e_k) - f(\theta))}{\epsilon} \eeq 
Therefore, 
\beq \frac{\partial (e_l \wedge \theta)}{\partial \theta_k} = \lim_{\epsilon \rightarrow 0} \frac{e_k \cdot (e_l \wedge (\theta + \epsilon e_k) - e_l \wedge \theta)}{\epsilon} = e_k \cdot (e_l \wedge e_k) = \nonumber \eeq
\beq = -e_k  \cdot (e_k \wedge e_l) = - e_k \cdot (e_k \cdot e_l (1- \delta^k_l)) = - (e_k \cdot e_k) \cdot e_l (1- \delta^k_l) = \nonumber \eeq 
\beq = - 1 \cdot e_l (1- \delta^k_l) = -e_l (1- \delta^k_l) \eeq 
Therefore, 
\beq \frac{\partial (\eta \wedge \theta)}{\partial \theta_k} = -\eta_{\perp k} \eeq
where $\eta_{\perp k}$ is defined as
\beq \eta = \sum_l x_l e_l \Longrightarrow \eta_{\perp k} = \sum_{l \neq k} x_l e_l \eeq 
In "usual situations" we have 
\beq \eta_{\perp k} \approx \eta \Longrightarrow \frac{\partial (\eta \wedge \theta)}{\partial \theta_k} \approx - \eta \eeq 
which is why we sloppily replace $\partial/ \partial \theta_k$ with $\partial/ \partial \theta$. However, if we consider non-analytic functions, things get a lot worse. For example, suppose
\beq f \bigg( \sum x_l e_l \bigg) = \sum x_{l+1} e_l \eeq 
then 
\beq \frac{\partial f}{\partial \theta_k} = \lim_{\epsilon \rightarrow 0} \frac{e_k \cdot (\epsilon e_{k-1})}{\epsilon} = e_k \cdot e_{k-1} = e_k \wedge e_{k-1} \eeq
which means that its dependence on the choice of $e_k$ is no longer negligible since it affects every single $k$ rather than just one of them. But, as long as we are dealing with the analytic functions, we will most likely approximate the conventional definition. 

\section{Conclusion}

In this paper we have shown that we can define the Grassmann integral as a limit of the sum, as opposed to merely an algebraic operation, if we obey the following conditions: 

1. Select a contour with $d$ orthogonal turns, each turn having the length of $a = \sqrt{2/d}$, where $d$ is a very large number. Admit that we would get an unwanted coefficient if said contour is rescaled

2. Use the LimMax instead of the ordinal limit in $d \rightarrow \infty$

3. Have two different products rather than just one.

Under those conditions, we have reproduced the conventional integral, up to sign disagreement. As explained in Section \ref{SignSection} said disagreement we introduced deliberately since we like our convention better, but it would take very little effort to go from our convention to the standard one, as described in Section \ref{SignSection}. 

In the process, we had to compute some of the "unusual" integrals, such as $(e_k * d \theta) \cdot \theta$. This, however, was necessary in order to arrive at the more conventional integrals: in the latter case, for example, it was needed in order to integrate $(d \theta_1 * d \theta_2) \cdot \theta_2$. In other words, we claim to reproduce all of the conventional results, with some "additional information" so to speak. 

Apart from that, we have found that we are able to integrate the non-analytic functions, in addition to integrating the analytic ones. As it stands, we haven't developed physical applications of the non-analytic functions. However, one idea that we might want to develop in the future is to invent a continuous measurement of the fermionic field (for example, use non-analytic Gaussians to write down the GRW collapse model for the fermionic field, which the analytic version of Gaussian won't fulfill since the analytic Gaussian of an anticommutting number is simply a constant, but the non-analytic doesn't have to be). As was stated in Conclusion of \cite{SverdlovBombelli}, such a model was previously impossible due to the fact that the Grassmann numbers don't have an ontological meaning, yet, again as suggested in \cite{SverdlovBombelli}, this situation has changed with the interpretation of the Grassmann numbers proposed in the current paper, which makes the idea of the continuous measurement of the fermionic field worth pursuing. Apart from the GRW model, we might also contemplate various Bohmian approaches with the fermionic field being used as beables.\footnote{But, not to confuse the reader, the specific non-analytic functions we have proposed in Sections \ref{NonAnalytic} and \ref{Derivatives} are useless as far as the above is concerned, they are only examples to draw home the concept that non-analytic functions are possible. As far as proposing the ones that might be useful for a measurement model, that is something for the future.}


\begin{thebibliography}{77}

\bibitem{SverdlovBombelli} R Sverdlov and L Bombelli, "Link between quantum measurement and the i$\epsilon$ term in the QFT propagator," arXiv:1306.1948, and Phys. Rev. D 90: 125020 (2014).

\bibitem{SpaceFillingCurve1} Cannon, James W.; Thurston, William P. (de) [1982], "Group invariant Peano curves", Geometry and Topology 11: 1315–1355, doi:10.2140/gt.2007.11.1315, ISSN 1465-3060, MR 2326947 

\bibitem{SpaceFillingCurve2} Mandelbrot, B. B. (1982), "Ch. 7: Harnessing the Peano Monster Curves", The Fractal Geometry of Nature, W. H. Freeman.

\bibitem{SpaceFillingCurve3} McKenna, Douglas M. (1994), "SquaRecurves, E-Tours, Eddies, and Frenzies: Basic Families of Peano Curves on the Square Grid", in Guy, Richard K.; Woodrow, Robert E., The Lighter Side of Mathematics: Proceedings of the Eugene Strens Memorial Conference on Recreational Mathematics and its History, Mathematical Association of America, pp. 49–73, ISBN 978-0-88385-516-4.

\bibitem{Mattress} Zee, "Quantum Field Theory in a Nutshell" Chapter I.3 "From Mattress to Field"

\bibitem{Cutoff} Zee, "Quantum Field Theory in a Nutshell" Chapter III.1 "Cutting off our ignorance", Subsection "parametrization of ignorance" p.146

\bibitem{Bohm1} D. D¨urr, S. Goldstein and N. Zangh`ı, “Quantum
equilibrium and the role of operators as observables
in quantum theory”, J. Stat. Phys. 116, 959-1055
(2004), and arXiv:quant-ph/0308038.

\bibitem{GRW1} G.C. Ghirardi, A. Rimini and T. Weber, “A model
for a unified quantum description of macroscopic
and microscopic systems”, in Quantum Probability
and Applications, L. Accardi et al. (eds), Springer,
Berlin, 1985.

\bibitem{GRW2} G.C. Ghirardi, A. Rimini and T. Weber, “Unified
dynamics for microscopic and macroscopic systems”,
Phys. Rev. D 34, 470 (1986).

\bibitem{Griffiths} David J. Griffiths, "Introduction to Quantum Mechanics", Second Edition, Copyright 2005 by Pearso Education, Inc. 

\bibitem{Grossberg} George B. Arfken, Hans J. Weber, "Mathematical Methods for Physicists", Academic Press, An Imprint of Elsevier. Copyright 2005, Elsevier Inc.

\end{thebibliography}
\end{document}